\documentclass[]{article}
\usepackage[utf8]{inputenc}
\usepackage[T1]{fontenc}
\usepackage[english]{babel}
\usepackage{amsmath}
\usepackage{amsthm, amssymb}
\usepackage{shadethm}
\usepackage{algpseudocode, algorithm, algorithmicx}
\usepackage{graphicx}
\usepackage{xcolor}
\usepackage{mathrsfs}
\usepackage{geometry}
\usepackage{enumerate}
\usepackage[colorlinks=true,linkcolor=blue]{hyperref}
\usepackage{mathtools}
\usepackage{thmtools}
\usepackage[nothm]{thmbox}
\usepackage{authblk}
\usepackage{comment}

\usepackage{subcaption}
\usepackage{soul}

\usepackage{array}
\newcolumntype{L}{>{$}l<{$}}

\newtheorem{theorem}{Theorem}[section]
\newtheorem{corollary}[theorem]{Corollary}

\newtheorem{lemma}[theorem]{Lemma}
\newtheorem{definition}[theorem]{Definition}
\newtheorem{remark}[theorem]{Remark}
\newenvironment{Proof}[1][]{\proof[#1]\normalsize}{\endproof}

\newcommand*\Let[2]{\State #1 $\gets$ #2}
\algrenewcommand\algorithmicrequire{\textbf{Given:}}
\algrenewcommand\algorithmicensure{\textbf{Return:}}

\renewcommand{\AA}{\mathbf{A}}
\newcommand{\HH}{\mathbf{H}}
\newcommand{\SIGMA}{\boldsymbol{\Sigma}}
\newcommand{\U}{\mathbf{U}}
\newcommand{\V}{\mathbf{V}}
\newcommand{\W}{\mathbf{W}}
\newcommand{\X}{\mathbf{X}}
\newcommand{\Z}{\mathbf{Z}}

\newcommand{\0}{\boldsymbol{0}}
\newcommand{\Alpha}{{\boldsymbol{\alpha}}}
\renewcommand{\a}{\mathbf{a}}
\newcommand{\Beta}{{\boldsymbol{\beta}}}
\newcommand{\Eta}{\boldsymbol{\eta}}
\newcommand{\g}{\mathbf{g}}

\newcommand{\THeta}{{\boldsymbol{\theta}}}
\renewcommand{\u}{\mathbf{u}}
\renewcommand{\v}{\mathbf{v}}
\newcommand{\w}{\mathbf{w}}
\newcommand{\x}{\mathbf{x}}
\newcommand{\XI}{\boldsymbol{\xi}}
\newcommand{\y}{\mathbf{y}}
\newcommand{\z}{\mathbf{z}}

\newcommand{\dist}{\mathrm{dist}}
\newcommand{\rank}{\mathrm{rank}}

\newcommand{\supp}{\mathrm{supp}}
\newcommand{\sign}{\mathrm{sign}}
\newcommand{\graph}{\mathrm{graph}}
\newcommand{\id}{\mathbf{Id}}

\newcommand{\dom}{\mathrm{dom}}
\newcommand{\trace}{\mathrm{tr}}
\newcommand{\A}{\mathcal{A}}

\renewcommand{\H}{\mathcal{H}}

\newcommand{\MM}{\mathbf{M}}

\newcommand{\R}{\mathbb{R}}

\newcommand{\abg}{{\Alpha, \Beta}}

\newcommand{\eps}{\varepsilon}

\newcommand{\tab}{\;\;\;\;}
\newcommand{\blue}[1]{#1}
\newcommand{\purple}[1]{#1}

\newcommand{\ol}[1]{\bar{#1}}
\renewcommand{\overline}[1]{\ol{#1}}
\newcommand{\vect}{\mathrm{vec}}

\newcommand{\prox}[1]{\mathrm{prox}_{#1}}

\newcommand{\E}[2][]{\;\mathbb{E}_{#1}\!\left[#2\right]}
\newcommand{\EE}{\mathbb{E}}

\renewcommand{\Pr}[2][]{\textnormal{\;\textsf{Pr}}_{#1}\!\left[#2\right]}

\newcommand{\round}[1]{\left( #1 \right)}

\newcommand{\curly}[1]{\left\{ #1 \right\}}

\newcommand{\abs}[2]{\left| #1 \right|_{#2}}
\newcommand{\inner}[2]{\left\langle #1 \right\rangle_{#2}}
\newcommand{\norm}[2]{\left\| #1 \right\|_{#2}}

\newcommand{\EN}{\mathrm{Enet}}

\newcommand{\rev}[1]{{#1}}

\DeclareMathOperator*{\argmin}{arg\,min}

\author[1,2]{Johannes Maly}
\affil[1]{\small Department of Mathematics, Ludwig-Maximilians-Universität, 80333 Munich, Germany}
\affil[2]{\small Munich Center for Machine Learning (MCML)}
\date{}
\begin{document}




\title{Robust Sensing of Low-Rank Matrices\\ with Non-Orthogonal Sparse Decomposition}
\maketitle

\begin{abstract}
We consider the problem of recovering an unknown low-rank matrix $\X_\star$ with (possibly) non-orthogonal, effectively sparse rank-$1$ decomposition from measurements $\y$ gathered in a linear measurement process $\A$. We propose a variational formulation that lends itself to alternating minimization and whose global minimizers provably approximate $\X_\star$ up to noise level. Working with a variant of robust injectivity, we derive reconstruction guarantees for various choices of $\A$ including sub-gaussian, Gaussian rank-$1$, and heavy-tailed measurements. Numerical experiments support the validity of our theoretical considerations.\\

\end{abstract}


\section{Introduction}

In this paper, we treat the reconstruction of sparse\footnote{\rev{A vector $\z \in \R^n$ is called $s$-sparse if at most $s$ of its entries are non-zero. On matrices sparsity can be defined and counted in various ways, cf.\ Section \ref{sec:Setting} below.}}, low-rank matrices $\X_\star \in \R^{n_1 \times n_2}$ from incomplete and inaccurate measurements $\y =  \mathcal A (\X_\star) + \Eta \in \R^m$, where $\A \colon \R^{n_1 \times n_2} \rightarrow \R^m$ resembles a linear measurement process and $\Eta \in \R^m$ models additive noise. This problem, which stems from compressed sensing \cite{foucart:2013} and related fields, is relevant in several modern applications such as sparse phase retrieval, blind deconvolution of sparse signals, machine learning, and data mining \cite{klibanov1995phase,iwen2014robust,haykin1994blind,lee2016blind,zou2006sparse,d2005direct}. 

\subsection{Related work I} 

Recovering low-rank matrices --- without additional sparsity constraints --- from linear measurements has been well-studied in the context of classical compressed sensing, i.e., compressed sensing of vectors \cite{candes2011tight, recht2010}. The bar is notably raised when the unknown matrix is assumed to be sparse and of low-rank, and both structures shall contribute in reducing the number of measurements $m$. As Oymak et al.\ pointed out in \cite{oymak2015simultaneously}, a mere linear combination of regularizers for different sparsity structures in general does not allow to outperform recovery guarantees of the ``best'' one of them alone. To further improve recovery, one has to go beyond linear combinations of already known convex regularizers. \\ 
A subtle approach to overcome the aforementioned limitations of purely convex methods is to assume a nested structure of the measurement operator $\A$ \cite{bahmani2016near,foucart2019jointly}. In this particular scenario, basic solvers for low-rank resp.\ row-sparse recovery can be applied in two consecutive steps. Although elegant, the nested approach clearly restricts possible choices for $\A$ and is of limited practical use.\\ 
In contrast, Lee et al.\ \cite{lee2013near} proposed and analyzed the so-called {\it Sparse Power Factorization} (SPF), a modified version of Power Factorization \cite{jain2013low}, without assuming any special structure of $\A$. Power Factorization recovers low-rank matrices by representing them as a product of two orthogonal matrices $\X = \U\V^\top$ and then applying alternating minimization over the (de)composing matrix $\U \in \mathbb{C}^{n_1\times R},\V \in \mathbb{C}^{n_2 \times R}$. To enforce sparsity of the columns of $\U$ and/or $\V$, SPF introduces Hard Thresholding Pursuit to each of the alternating steps. Lee et al.\ were able to show that, using suitable initializations and assuming the noise level to be sufficiently small, SPF approximates low-rank matrices $\X$ that are row- and/or column-sparse from a nearly optimal number of measurements: If $\X$ is rank-$R$, has $s_1$-sparse columns and $s_2$-sparse rows, then $m \gtrsim R(s_1 + s_2) \log (\max \{en_1/s_1, en_2/s_2\})$ Gaussian measurements suffice for robust recovery,  which is up to the log-factor at the information theoretical limit. Despite its theoretical optimality, the setting of SPF is actually quite restrictive as all columns (resp.\ rows) need to share a common support and the matrices $\U,\V$ need to be orthogonal. On the one hand, empirically, it has been shown in  \cite{lee2013near} that SPF outperforms methods based on convex relaxation. On the other hand, SPF is heavily based on the assumption that the operator $\mathcal A$ possesses a suitable restricted isometry property and cannot be applied to arbitrary inverse problems, as it may even fail to converge otherwise. The reason is that SPF is based on hard-thresholding  \cite{BLUMENSATH2009265}, which is not a Lipschitz continuous map. Let us mention that in \cite{lee2016blind} the authors extended their analysis of SPF to the measurement set-up of blind deconvolution.\\
Inspired by recent works on multi-penalty regularization \cite{naumova2014minimization, naumova2016}, the authors of \cite{fornasier2018robust} aimed at enhancing robustness of recovery by alternating minimization of an $\ell_1$-norm based multi-penalty functional. Though not as close to the information theoretical limit as SPF, the theoretical results therein hold for arbitrarily large noise magnitudes and a wider class of ground-truth matrices than the one considered in \cite{lee2013near}.\\
Let us finally point out that a closely related line of work comes from statistical literature under the name \emph{Sparse Principal Component Analysis} (SPCA) \cite{zou2006sparse,d2005direct}. In order to defeat the curse of dimensionality when finding principal subspaces of covariance matrices, SPCA admits non-orthogonal subspace decompositions and enforces sparsity on the vectors spanning the respective spaces. However, observations in SPCA are provided from noisy samples of the underlying distribution, whereas in our case the matrix itself is observed indirectly. It is thus hard to directly compare results of this paper with corresponding results for SPCA.

\subsection{Problem setting}
\label{sec:Setting}

Let us begin by specifying the problem at hand. Given a linear measurement operator $\A \colon \R^{n_1 \times n_2} \rightarrow \R^m$ and a corrupted vector of measurements
\begin{align} \label{A}
    \y = \A (\X_\star) + \Eta = {\frac{1}{\sqrt m}}\begin{pmatrix}
    \langle \AA_1,\X_\star \rangle_F \\
    \vdots \\
    \langle \AA_m,\X_\star \rangle_F	
    \end{pmatrix} + \Eta,
\end{align}
we wish to estimate the unknown signal matrix $\X_\star \in \R^{n_1\times n_2}$. Here, the vector $\Eta \in \R^m$, of which only the $\ell_2$-norm is (approximately) known, models additive noise. 
Note that $\A$ is fully determined by the $m$ matrices $\AA_i \in \R^{n_1 \times n_2}$ and that individual measurements correspond to the Frobenius products $\langle \AA_i,\X_\star \rangle_F = \trace(\AA_i \X_\star^\top)$, for $i \in [m] := \curly{1,\dots,m}$.

Whereas \eqref{A} is ill-posed if $m < n_1 n_2$, it becomes well-posed if we assume some prior knowledge on $\X_\star$. 
\rev{In the rest of the work, we thus suppose that $\X_\star$ is of rank $R \ge 1$ and possesses a decomposition of the form
\begin{align} \label{xSVD}
    \X_\star = \U_\star \V_\star^\top, 
\end{align}
where $\U_\star \in \R^{n_1 \times R}$ and $\V_\star \in \R^{n_2\times R}$ are (effectively) sparse. Effective sparsity is a generalized notion of sparsity introduced by Plan and Vershynin in \cite[Section 3]{Plan2013LP}. 
\begin{definition}[Effectively sparse vectors] \label{def:EffectivelySparse}
	The set of effectively $s$-sparse vectors of dimension $n$ is defined by 
	\begin{align*}
	K_{n,s} = \{ \z \in \R^n \colon \Vert \z \Vert_2 \le 1 \text{ and } \Vert \z \Vert_1 \le \sqrt{s} \}.
	\end{align*}
\end{definition}
\begin{remark}
	Note that any unit-norm $s$-sparse vector is also effectively $s$-sparse. Effectively sparse vectors are well-approximated by sparse vectors as made precise in \cite[Lemma 3.2]{Plan2013LP}. Moreover, the set $K_{n,s}$ may be viewed as a convex hull of the set of $s$-sparse unit-norm vectors \cite[Lemma 3.1]{Plan2013LP}.
\end{remark} 

We call a decomposition as in \eqref{xSVD} \emph{sparse decomposition (SD) of $\X_\star$} and say $\U_\star$ and $\V_\star$ are the left resp.\ right SD component. If the columns of $\U_\star$ and $\V_\star$ are orthogonal, then we call \eqref{xSVD} \emph{orthogonal SD of $\X_\star$}. In particular, we consider two specific sets of matrices having an SD: 

\paragraph{(I)} The set of rank-$R$ matrices whose columns and rows are jointly $s_1$- resp.\ $s_2$-sparse
\begin{align} \label{eq:Sbar}
    \bar{S}_{s_1,s_2}^R
    = \{ \Z \in \R^{n_1 \times n_2} \colon \rank(\Z) \le R, \; \| \Z \|_{2,0} \le s_1, \text{ and } \| \Z \|_{0,2} \le s_2  \},
\end{align}
where $\|\cdot\|_{2,0}$ and $\|\cdot\|_{0,2}$ count the number of non-zero columns and rows. If $\X_\star \in \bar{S}_{s_1,s_2}^R$ with singular value decomposition (SVD)
\begin{align} \label{SVD}
    \X_\star = \U\SIGMA \V^\top = \sum_{r = 1}^{R} \sigma_r \u^r (\v^r)^\top,
\end{align}
where $\SIGMA$ is a diagonal matrix containing the singular values $\sigma_1 \ge ... \ge \sigma_{R} > 0$ and $\U \in \R^{n_1 \times R}$ and $\V \in \R^{n_2 \times R}$ have orthonormal columns which are called left and right singular vectors, then $\U_\star = \U \SIGMA^{\frac{1}{2}}$ and $\V_\star = \V \SIGMA^{\frac{1}{2}}$ form an orthogonal SD of $\X_\star$.

\paragraph{(II)} The set of rank-$R$ matrices that have an SD whose left and right SD components are \emph{effectively} $Rs_1$- and $Rs_2$-sparse 
\begin{align*}
	K_{s_1,s_2}^{R} = \curly{ \Z = \U\V^\top \colon \U \in \R^{n_1 \times R}, \; \V \in \R^{n_2 \times R} \text{ and } \vect(\U) \in \sqrt{R} K_{R n_1, R s_1}, \; \vect(\V) \in \sqrt{R} K_{R n_2, R s_2}  },
\end{align*}
where $\vect(\cdot)$ denotes the vectorization of a matrix.
The set $K_{s_1,s_2}^{R}$ can be viewed as a natural relaxation of the set of rank-$R$ matrices that have an SD whose left and right SD components are $Rs_1$- and $Rs_2$-sparse
\begin{align} \label{eq:S}
    S_{s_1,s_2}^R
    = \{ \Z = \U\V^\top \colon \U \in \R^{n_1 \times R}, \; \V \in \R^{n_2 \times R} \text{ and } \| \U \|_{0} \le Rs_1, \text{ and } \| \V \|_{0} \le Rs_2  \},
\end{align}
where $\| \cdot \|_0$ counts the number of non-zero entries of a vector/matrix. The difference between \eqref{eq:Sbar} and \eqref{eq:S} is that the former requires the SD components $\U$ and $\V$ to have jointly $s_1$- resp.\ $s_2$-sparse columns whereas the latter allows an arbitrary distribution of the non-zero entries over the columns of $\U$ and $\V$. Note that $\Z \in \bar{S}_{s_1,s_2}^R$ implies $\Z \in S_{s_1,s_2}^R$ and that any $\Z \in S_{s_1,s_2}^R$ that has an SD whose left and right components have jointly sparse columns lies in $\bar{S}_{s_1,s_2}^R$.
Furthermore, any $\Z \in S_{s_1,s_2}^R$ satisfies $\Z \in \bar S_{Rs_1,Rs_2}^R$.

By appealing to the definition of $K_{n,s}$ in Definition \ref{def:EffectivelySparse}, the set $K_{s_1,s_2}^R$ can equivalently be written as
\begin{align} \label{eq:K}
	K_{s_1,s_2}^{R} = \curly{ \Z = \U\V^\top \colon \max \curly{ \norm{\U}{F}^2, \norm{\V}{F}^2} \le R \text{ and } \norm{\U}{1} \le R \sqrt{s_1}, \; \norm{\V}{1} \le R \sqrt{s_2} },
\end{align}
where $\norm{\cdot}{1}$ denotes the vector $\ell_1$-norm and is, by abuse of notation, applied to matrices, i.e., $\norm{\Z}{1} = \sum_{i,j} \abs{Z_{i,j}}{}$ for any matrix $\Z$. If the SVD of a matrix $\Z$ satisfies $\max \{ \| \U\SIGMA^{\frac{1}{2}} \|_{F}^2, \| \V\SIGMA^{\frac{1}{2}} \|_{F}^2 \} \le R$, $\| \U\SIGMA^{\frac{1}{2}} \|_{1} \le R \sqrt{s_1}$, and $\| \V\SIGMA^{\frac{1}{2}} \|_{1} \le R \sqrt{s_2}$, then
\begin{align*}
    \Z = (\U\SIGMA^{\frac{1}{2}}) (\V \SIGMA^{\frac{1}{2}})^\top
\end{align*}
is an orthogonal SD and $\Z \in K_{s_1,s_2}^R$. 
In particular, the set $K_{s_1,s_2}^R$ includes all matrices $\Z \in S_{s_1,s_2}^R$ that have an SD whose left and right SD components have columns with Euclidean norm bounded by one.

\paragraph{} 
The set $\bar{S}_{s_1,s_2}^R$, which comes with the strongest assumptions, has been considered in many previous works \cite{lee2013near,foucart2019jointly,eisenmann2021riemannian}.
Indeed, whereas the orthogonal SDs of matrices in $\bar{S}_{s_1,s_2}^R$ are closely tied to the SVD and thus unique up to scaling ambiguities and row/column permutations (if $\Z = \U\V^\top$ is an orthogonal SD, then $\Z = (\U\mathbf D)(\V\mathbf D^{-1})^\top$ is also an orthogonal SD, for any invertible diagonal matrix $\mathbf D \in \R^{R\times R}$),
SDs of matrices in $S_{s_1,s_2}^R$ and $K_{s_1,s_2}^R$ are non-unique in general due to the possible non-orthogonality of the component columns. What is more, the set $K_{s_1,s_2}^R$ is not even scaling invariant, i.e., $\Z \in K_{s_1,s_2}^R$ does not imply $\gamma\Z \in K_{s_1,s_2}^R$, for $\gamma \in \R$. 

\begin{remark}
\label{rem:EffectiveSparsity}
    A scaling-invariant definition of effective sparsity that is related to Definition \ref{def:EffectivelySparse} is to call $\z$ effectively $s$-sparse if $\tfrac{\| \z \|_1^2}{\| \z \|_2^2}~\le~s$, i.e., if $\tfrac{\z}{\| \z \|_2} \in K_{n,s}$. Since the scaling-invariance of the sparsity model does not imply scaling-invariance of our matrix signal model in \eqref{eq:K}, cf.\ Remark \ref{DominiksRemark} below, and since Definition \ref{def:EffectivelySparse} naturally aligns with our reconstruction method described in Section \ref{sec:Contribution} below, also see Lemma \ref{Bound-uv2}, we use the former definition despite its scaling variance.
\end{remark}

One important feature that however all presented sets share is that, to a certain extent, they are closed under summation.

\begin{lemma} \label{lemma:sumrule}
    Let $R, s_1, s_2 \in \mathbb N$ and $\Gamma > 0$ be fixed.
    \begin{enumerate}[(i)]
        \item If $\Z,\Z' \in \bar{S}_{s_1,s_2}^R$, then $\Z + \Z' \in \bar{S}_{2s_1,2s_2}^{2R}$
        \item If $\Z,\Z' \in S_{s_1,s_2}^R$, then $\Z + \Z' \in S_{s_1,s_2}^{2R}$
        \item If $\Z,\Z' \in \Gamma K_{s_1,s_2}^R$, then $\Z + \Z' \in \Gamma K_{s_1,s_2}^{2R}$.
    \end{enumerate}
\end{lemma}

To simplify notation, we will often assume $s_1 = s_2 = s$ in the following and use the shorthand notations $\bar{S}_s^R = \bar{S}_{s,s}^R$, $S_s^R = S_{s,s}^R$, and $K_s^R = K_{s,s}^R$. It is straight-forward to generalize respective results to the case $s_1 \neq s_2$.
}

\subsection{Contribution} 
\label{sec:Contribution}

We build upon ideas in \cite{fornasier2018robust} and reconstruct $\X_\star$ by a variational approach. It is based on alternating minimization of the multi-penalty functional $J_\abg^R \colon \R^{n_1 \times R} \times \R^{n_2 \times R} \rightarrow \R$ defined, for $\alpha_1,\alpha_2, \beta_1, \beta_2 > 0$, by
\begin{equation} \label{Jab0}
    J_{\Alpha,\Beta}^R (\U,\V) := \left\Vert \y - \mathcal A \left( \U\V^\top \right) \right\Vert_2^2 + \alpha_1 \norm{\U}{F}^2 + \alpha_2 \norm{\U}{1} + \beta_1 \norm{\V}{F}^2 + \beta_2 \norm{\V}{1}.
\end{equation}
where $\Alpha = (\alpha_1, \alpha_2)^\top, \Beta = (\beta_1, \beta_2)^\top$ are regularization parameters and $\norm{\cdot}{1}$ denotes the vector $\ell_1$-norm. The functional in \eqref{Jab0} can be interpreted as a generalization of the linear regression based SPCA approach in \cite{zou2006sparse}.  
Despite the convex multi-penalty regularization term $\alpha_1 \norm{\U}{F}^2 + \alpha_2 \norm{\U}{1} + \beta_1 \norm{\V}{F}^2 + \beta_2 \norm{\V}{1}$, also known as elastic net \cite{zou2005regularization}, the functional \eqref{Jab0} is highly non-convex; hence, in light of the negative results in \cite{oymak2015simultaneously} it provides hope for better performance than lifting and convex relaxation. At the same time, one notices that $J_\abg^R$ becomes convex when $\U$ or $\V$ are fixed. We can thus efficiently minimize \eqref{Jab0} by  alternating schemes, e.g., alternating minimization 
\begin{equation}
	 \label{ATLAS0}
    \begin{cases}
    \U_{k+1} 
    &=  \argmin_\U \left\Vert \y - \A(\U \V_k^\top) \right\Vert^2_2 + \alpha_1 \Vert \U \Vert_F^2 + \alpha_2 \norm{\U}{1} \\ 
    \V_{k+1} 
    &=  \argmin_\V \left\Vert \y - \A(\U_{k+1} \V^\top) \right\Vert_2^2 + \beta_1 \Vert \V \Vert_F^2 + \beta_2 \norm{\V}{1}. \\ 
\end{cases}
\end{equation}
\rev{The fact that minimizers of $J_\abg^R$ do not consist of orthogonal matrices in general, excludes the use of \eqref{eq:Sbar} in the analysis and motivates the use of \eqref{eq:S} resp.\ \eqref{eq:K} despite their challenging non-orthogonal decomposition structure; at least if one aims for tight bounds (recall from Section \ref{sec:Setting} that $\Z \in S_{s_1,s_2}^R$ only implies $\Z \in \bar{S}_{Rs_1,Rs_2}^R$ and \emph{not} $\Z \in \bar{S}_{s_1,s_2}^R$, i.e., considering non-orthogonal minimizers as elements of $\bar{S}_{Rs_1,Rs_2}^R$ in the analysis would add an artificial $R$ dependence). At the same time, the more general signal set $K_{s_1,s_2}^R$ in \eqref{eq:K} allows to cover interesting data that is not exactly $s$-sparse, cf.\ Section \ref{Numerics3}.

In addition to proposing the functional $J_\abg^R$ and the signal set $K_{s_1,s_2}^R$ for its analysis, our main contribution is threefold:
\begin{enumerate}
    \item Under minimal assumptions on $\A$, we show for any global minimizer $(\U_\abg,\V_\abg)$ of \eqref{Jab0}: if $\X_\star \in \Gamma K_s^R$, for $\Gamma > 0$, and $(\abg)$ are suitably chosen, then $\X_\abg = \U_\abg \V_\abg^\top \in \Gamma K_s^{4R}$ and $\A(\X_\abg) \approx \y$, i.e., $\X_\abg$ is regular and satisfies the measurements up to noise level (Lemmas \ref{Bound-y2} and \ref{Bound-uv2}).
    \item If $\A$, in addition, satisfies a robust injectivity property on $\Gamma K_s^{5R}$ (Definition \ref{apprRIPDef}), then any $\X_\abg$ as defined in the first bullet point satisfies $\X_\abg \approx \X_\star$ up to noise level and an injectivity constant $\delta$ (Theorem \ref{ApproxX}).
    \item Various types of measurement operators $\A$ (subgaussian, heavy-tailed, rank-$1$) satisfy the required robust injectivity (Theorem \ref{thm:SmallBallRIP} and Corollary \ref{GaussianRIP}).
\end{enumerate}
Furthermore, we examine two approaches to minimizing \eqref{Jab0}, for which we show global convergence to stationary points and local convergence to global minimizers, cf.\ Section \ref{sec:Convergence}.

\textbf{Our contribution in light of \cite{fornasier2018robust}:} Although the functional in \eqref{Jab0} and the signal set $\Gamma K_{s_1,s_2}^R$ in \eqref{eq:K} appear to be conceptually close to their predecessors in \cite{fornasier2018robust}, this work notably contributes to the existing theory. \\
First, the use of elastic net as sparsity regularizer in \eqref{Jab0} allows to control both $\ell_1$- and $\ell_2$-norm simultaneously and leads to improved regularity of minimizers, cf. Lemma \ref{Bound-uv2} and the consequent discussion. \\
Second, the definition of $\Gamma K_{s_1,s_2}^R$ treats the component matrices $\U$ and $\V$ as a whole such that their non-zero entries may be distributed without further restrictions. This is crucial to derive the regularity in Lemma \ref{Bound-uv2} and stands in contrast to \cite{fornasier2018robust} where the sparsity of each column of $\U_\abg$ and $\V_\abg$ needed to be controlled individually. Together with the first-mentioned point this leads to a cleaner analysis of the approximation error and removes artificial technical assumptions on the computed minimizer $(\U_\abg,\V_\abg)$, see \cite[Corollary 4.4]{fornasier2018robust} in comparison to Theorem \ref{ApproxX} below. What is more, in Theorem \ref{ApproxX} the parameters $\abg$ may be chosen arbitrarily close to zero without worsening the regularity of $\X_\abg$. This is required for covering low-noise regimes but stands in stark contrast to \cite[Theorem 4.3]{fornasier2018robust}. \\
Third, in contrast to \cite[Definition 4.1]{fornasier2018robust} (and other previous works like \cite{lee2013near}) the generalized robust injectivity we define in Definition \ref{apprRIPDef} is not restricted to measurement operators with strongly concentrating distributions, cf.\ Theorem \ref{thm:SmallBallRIP}. When applied to subgaussian measurements, Theorem \ref{thm:SmallBallRIP} even reduces the sufficient sample complexity in \cite[Lemma 5.3]{fornasier2018robust} by a factor $\Gamma^2$ and, for unit norm $\X_\star \in \bar{S}_{s_1,s_2}^R$, reaches the near-optimal sample complexity in \cite{lee2013near} (up to log-factors), cf.\ Corollary \ref{GaussianRIP} and Remark \ref{rem:dim}.  \\
Finally, our local convergence analysis repairs a theoretical inaccuracy of \cite{fornasier2018robust}. Indeed, in \cite{fornasier2018robust} the local convergence analysis considers Proximal Alternating Minimization, whereas the algorithm ATLAS \cite[Algorithm 1]{fornasier2018robust} is an alternating minimization like Algorithm \ref{alg:AM} below. To close this gap between empirical evaluation and theory, we analyze both Proximal Alternating Linearized Minimization and Alternating Minimization in our setting. 
}

Like with most of the non-convex methods, convergence to global minimizers and thus empirical performance of \eqref{ATLAS0} depends on a proper initialization $(\U_0, \V_0)$. We, however, do not provide such an initialization method but leave it as an open problem for future research. Our empirical evaluation in Section \ref{Numerics} suggests that the for low-rank recovery commonly used spectral initialization leads to decent results.

\subsection{Related Work II}

Let us briefly comment on the very last point. Finding a reliable and tractable initialization procedure is challenging. This reflects the intrinsic hardness of the presented problem, a hardness that stems from its deep connection to SPCA, which is known to be NP-hard in general \cite{magdon2017np}. To the best of our knowledge, all yet existing approaches to the reconstruction of jointly sparse and low-rank matrices share this impediment in one form or another. The above mentioned guarantees for SPF, which come with a tractable initialization, are restricted to signals with few dominant entries \cite{lee2013near,geppert2019sparse}. The recent work \cite{eisenmann2021riemannian}, which builds upon ideas on generalized projections in \cite{foucart2019jointly} and uses a Riemannian version of Iterative Hard Thresholding to reconstruct jointly row-sparse and low-rank matrices, only provides local convergence results. The alternative approach of using optimally weighted sums or maxima of convex regularizers \cite{kliesch2019simultaneous} --- the only existing work apart from our predecessor paper \cite{fornasier2018robust} that considers non-orthogonal sparse low-rank decompositions --- requires optimal tuning of the parameters under knowledge of the ground-truth. Whereas one may not hope for a general solution to these problems, the previously mentioned tractable results on nested measurements \cite{bahmani2016near,foucart2019jointly} suggest that suitable initialization methods can be found for specific applications. 

\subsection{Outline and Notation} 

The organization of the paper is as follows. Section \ref{MainResults} provides the main results and part of the proofs. The remaining proofs can be found in Section \ref{Proofs}. In Section \ref{Numerics}, we compare our theoretical findings to actual empirical evidence. We conclude in Section \ref{Conclusion} with a discussion on open problems and future work.\\

We use the shorthand notation $[n] := \{ 1,...,n \}$ to write index sets. 
The relation $a \gtrsim b$ is used to express $a \ge Cb$ for some positive constant $C$, and $a \simeq b$ stands for {$a \gtrsim b$} and {$b \gtrsim a$}.\\
For a matrix $\Z \in \R^{n_1 \times n_2},$ we denote its  transpose by $\Z^\top$. The support of $\Z$, i.e., the index set of the non-zero entries, is denoted by $\supp(\Z)$. The function $\vect$ vectorizes any matrix and $\vect^{-1}$ reverses the vectorization. Hence, $\vect(\Z) \in \R^{n_1 n_2}$ and $\Z = \vect^{-1}(\vect(\Z))$. We denote by $\Vert \Z \Vert_1$ the $\ell_1$-norm of $\vect(\Z)$, by $\Vert \Z \Vert_F$ the Frobenius norm of $\Z$ ($\ell_2$-norm of the vector of singular values), by $\Vert \Z \Vert_{2 \rightarrow 2}$ the operator norm of $\Z$ (top singular value), and by $\norm{\Z}{*}$ the nuclear norm of $\Z$ (sum of singular values). \\
The set-valued operator $\partial$ denotes the limiting Fr\'echet subdifferential, and $\operatorname{dom} \partial f = \curly{\x \in \R^n \colon \partial f (\x) \neq \emptyset}$ its domain when applied to a function $f \colon \mathbb{R}^n \rightarrow \mathbb{R} \cup \curly{\infty}$, cf.\ \cite{rockafellar2009variational,mordukhovich2006variational}. \\
The covering number $N(M,\Vert \cdot \Vert,\eps)$ of a set $M$ is the minimal number of $\Vert \cdot \Vert$-balls of radius $\eps$ that are needed to cover the set $M$. The cardinality of any $\eps$-net $\tilde{M}$ of $M$, i.e., for all $\z \in M$ there is $\tilde{\z} \in \tilde{M}$ with $\Vert \z - \tilde{\z} \Vert < \eps$, yields an upper bound for $N(M,\Vert \cdot \Vert,\eps)$.


\section{Main Results} \label{MainResults}

We are now ready to state the main results of the paper. 
In Section \ref{sec:Main1}, we show how minimizers of $J_\abg^R$ yield under minimal assumptions solutions to the inverse problem \eqref{A}. In Section \ref{sec:RecoveryPropertiesRIP}, we estimate the approximation error assuming robust injectivity of $\A$ on $K_{s_1,s_2}^R$. We then present in Section \ref{SectionRIP} various types of operators fulfilling robust injectivity and, finally, provide local convergence guarantees for \eqref{ATLAS0} and related methods in Section \ref{sec:Convergence}.


\subsection{General Properties of Minimizers} \label{sec:Main1}

\rev{Let us begin with the basic properties that all minimizers of $J_\abg^R$ have under very general assumptions. To this end, let
\begin{align} \label{eq:Jmin}
    (\U_\abg, \V_\abg) \in \argmin_{(\U,\V) \in \R^{n_1\times R} \times \R^{n_2\times R}} J_\abg^R (\U,\V)
\end{align}
be any minimizer of $J_\abg^R$ and denote
\begin{align} \label{Xab}
    \X_\abg = \U_\abg \V_\abg^\top.
\end{align} 
The first result bounds the measurement misfit of any such $\X_\abg$. The result is a straight-forward modification of \cite[Proposition 3.1]{fornasier2018robust}.}
\begin{lemma}[Measurement misfit] \label{Bound-y2}
	Assume $\X_\star$ with $\rank(\X) \le R$ is generating the noisy measurements $\y~=~\A(\X_\star)~+~\Eta$ and let $(\U_\abg, \V_\abg)$ be a global minimizer of $J_\abg^R$ where $\Alpha$ and $\Beta$ are chosen sufficiently small (depending on $\| \X_\star \|_{F}$ and $\norm{\Eta}{2}$). Then,
	\begin{align}\label{measfit}
	\Vert \y - \A(\X_\abg) \Vert_2 \le 2 \Vert \Eta \Vert_2.
	\end{align}
\end{lemma}
\begin{Proof}
    \rev{
	Let $\X_\star = \U\SIGMA \V^\top$ denote the SVD of $\X_\star$ where $\U \in \R^{n_1 \times R}$ and $\V \in \R^{n_2 \times R}$. Since $(\U_\abg,\V_\abg)$ minimizes $J_\abg^R$, we may use $(\U\SIGMA^\frac{1}{2}, \SIGMA^\frac{1}{2} \V)$ as a competitor in \eqref{eq:Jmin} to obtain
	\begin{align*}
		\norm{\y - \A(\X_\abg)}{2}^2 
		&\le J_\abg^R(\U_\abg,\V_\abg) 
		\le J_\abg^R (\U\SIGMA^\frac{1}{2}, \SIGMA^\frac{1}{2} \V) \\
		&= \norm{\Eta}{2}^2 + \alpha_1 \norm{\U\SIGMA^\frac{1}{2}}{F}^2 + \alpha_2 \norm{\U\SIGMA^\frac{1}{2}}{1} + \beta_1 \norm{\SIGMA^\frac{1}{2} \V}{F}^2 + \beta_2 \norm{\SIGMA^\frac{1}{2} \V}{1} \\
		&\le 4 \norm{\Eta}{2}^2,
	\end{align*} 
	for $\alpha_1 \le  \frac{\norm{\Eta}{2}^2}{2 \norm{\U\SIGMA^\frac{1}{2}}{F}^2}$, $\alpha_2 \le  \frac{\norm{\Eta}{2}^2}{2 \norm{\U\SIGMA^\frac{1}{2}}{1}}$, $\beta_1 \le  \frac{\norm{\Eta}{2}^2}{2 \norm{\SIGMA^\frac{1}{2} \V}{F}^2}$, and $\beta_2 \le  \frac{\norm{\Eta}{2}^2}{2 \norm{\SIGMA^\frac{1}{2} \V}{1}}$.
	}
\end{Proof}
The second result states that if the ratio of the parameters is fixed in a suitable way and $\X_\star$ is low-rank with effectively sparse decomposition, then the same holds true for global minimizers of $J_\abg^R$ as long as $\alpha$ and $\beta$ are chosen sufficiently large to avoid overfitting.
\begin{lemma}[Regularity] \label{Bound-uv2}
	Assume $\X_\star \in \Gamma K_s^R$, for $\Gamma > 0$, is generating the noisy measurements $\y~=~\A(\X_\star)~+~\Eta$ and let $(\U_\abg, \V_\abg)$ be a global minimizer of $J_\abg^R$ where $\alpha_1 = \sqrt{\frac{s}{\Gamma}} \alpha_2 = \beta_1 = \sqrt{\frac{s}{\Gamma}} \beta_2$. If $\| \y - \A(\X_\abg) \|_2~\ge~\| \Eta \|_2$, we have that $\X_\abg \in \Gamma K_s^{4R}$.
\end{lemma}
\begin{Proof}
	Let $\X_\star = \U_\star \V_\star^\top$ denote an SD of $\X_\star \in \Gamma K_s^R$ such that
	\begin{align*}
	    \max \curly{\| \U_\star\|_{F}^2, \| \V_\star \|_{F}^2} \le \Gamma R
	    \quad \text{ and } \quad
	    \max \curly{\| \U_\star\|_{1}, \| \V_\star \|_{1}} \le R \sqrt{\Gamma s}
	\end{align*} 
	(note that such an SD always exists by taking an arbitrary SD of $\frac{1}{\Gamma} \X_\star \in K_s^R$ and multiplying both the left and right components with $\sqrt{\Gamma}$). By minimality of $\X_\abg$, we get that
	\begin{align*}
		&\norm{\y - \A(\X_\abg)}{2}^2 + \alpha_1 \norm{\U_\abg}{F}^2 + \alpha_2 \norm{\U_\abg}{1} + \beta_1 \norm{\V_\abg}{F}^2 + \beta_2 \norm{\V_\abg}{1} \\
		&= J_\abg^R (\U_\abg, \V_\abg) \\
		&\le J_\abg^R (\U_\star, \V_\star) \\
		&= \norm{\Eta}{2}^2 + \alpha_1 \| \U_\star \|_{F}^2 + \alpha_2 \| \U_\star\|_{1} + \beta_1 \| \V_\star \|_{F}^2 + \beta_2 \| \V_\star \|_{1}.
	\end{align*}
	Subtracting $\norm{\y - \A(\X_\abg)}{2}^2$ on both sides and using that by assumption $\norm{\Eta}{2} \le \norm{\y - \A(\X_\abg)}{2}$ and $\alpha_1 = \sqrt{\frac{s}{\Gamma}} \alpha_2 = \beta_1 = \sqrt{\frac{s}{\Gamma}} \beta_2$ leads to
	\begin{align*}
		\max \curly{\norm{\U_\abg}{F}^2, \norm{\V_\abg}{F}^2}
		&\le \| \U_\star \|_{F}^2 + \sqrt{\frac{\Gamma}{s}} \| \U_\star \|_{1} + \| \V_\star \|_{F}^2 + \sqrt{\frac{\Gamma}{s}} \| \V_\star \|_{1} \\
		&\le 4 R \Gamma
	\end{align*}
	and
	\begin{align*}
		\max \curly{\norm{\U_\abg}{1}, \norm{\V_\abg}{1}}
		&\le \sqrt{\frac{s}{\Gamma }} \| \U_\star\|_{F}^2 + \|\U_\star \|_{1} + \sqrt{\frac{s}{\Gamma }} \| \V_\star \|_{F}^2 + \| \V_\star \|_{1} \\
		&\le 4 R \sqrt{\Gamma s},
	\end{align*}
	which shows that $\X_\abg \in \Gamma K_s^{4R}$.
\end{Proof}
The assumption $\| \y - \A(\X_\abg) \|_2 \ge \| \Eta \|_2$ in Lemma \ref{Bound-uv2} is not restrictive. As soon as $\| \y - \A(\X_\abg) \|_2 = \| \Eta \|_2$, decreasing $\Alpha$ and $\Beta$ any further becomes undesirable since this will lead to overfitting. \rev{Comparing Lemma \ref{Bound-uv2} to \cite[Lemma 3.3]{fornasier2018robust}, we see a massive improvement: First, by exploiting the Frobenius-norm control of the elastic-net regularizers, Lemma \ref{Bound-uv2} contains no implicit assumptions on the specific shape of $\X_\abg$ or its SD. Second, by the refined definition of the signal set (compare $\Gamma K_{s_1,s_2}^{R}$ in \eqref{eq:K} to $K_{s_1,s_2}^{R,\Gamma}$ in \cite{fornasier2018robust}), the lemma guarantees that $\X_\abg$ has the same regularity as $\X_\star$ instead of only bounding the effective sparsity of single columns of $\U_\abg$ and $\V_\abg$. }

All in all, the results of Section \ref{sec:Main1} show that, for any $\A$ and well-tuned parameters $\Alpha$ and $\Beta$, any $\X_\abg$ is a reasonable approximation of $\X_\star \in \Gamma K_s^R$. Indeed, it is of rank $R$, fulfills the measurements up to noise level, and has an effectively sparse decomposition of the same order as $\X_\star$. However, the parameters $\Alpha$ and $\Beta$ have to be chosen with care, neither too small nor too large. Moreover, Lemma \ref{Bound-uv2} shows that $\Alpha$ and $\Beta$ should be of similar magnitude. Otherwise either the left or the right components of $\X_\abg$ cannot be controlled. For further information on the parameter tuning we refer the reader to Section \ref{Numerics} and the discussion following Theorem \ref{ApproxX}.


\subsection{Reconstruction Properties of Minimizers} 
\label{sec:RecoveryPropertiesRIP}

To obtain proper reconstruction guarantees, we introduce a robust variant of injectivity for linear operators $\A$ acting on $K_{s_1,s_2}^R$. These properties may be viewed as generalizations of the rank-$R$ and $(s_1,s_2)$-sparse RIP of Lee et.\ al.\ in  \cite{lee2013near}. 
\begin{definition}[Robust injectivity] \label{apprRIPDef}
	Let $\Gamma > 0$. We say that $\A$ satisfies robust injectivity on $\Gamma K_{s_1,s_2}^R$ with injectivity constants $\gamma \in (0,1]$ and $\delta > 0$ if
	\begin{align} \label{apprRIP}
	\norm{\A(\Z)}{2}^2 \ge \gamma \norm{\Z}{F}^2 - \delta
	\end{align}
	for all $\Z \in \Gamma K_{s_1,s_2}^{R}$.
\end{definition}
The requirement in \eqref{apprRIP} is weak when compared to classical restricted isometry properties \cite{foucart:2013}. It only lower bounds the contractivity of $\A$ when applied to matrices in $\Gamma K_{s_1,s_2}^R$. \rev{Furthermore, since the set $\Gamma K_{s_1,s_2}^R$ is not scaling invariant, the corresponding injectivity property in \eqref{apprRIP} is neither, see also Remark \ref{DominiksRemark} below.} \\
The main reconstruction result reads as follows: If $\A$ is injective in the sense of Definition \ref{apprRIPDef} and $\Alpha, \Beta$ are suitably chosen, any global minimizer of $J_\abg^R$ provides an approximation of $\X_\star$ up to noise level and injectivity constant $\delta$. Recall from above that w.l.o.g.\ and for the sake of simplicity we restrict ourselves to the case $s_1 = s_2 = s$ and $K_s^R = K_{s,s}^R$ here.
\begin{theorem}[Reconstruction of signals] \label{ApproxX}
    Let $\Gamma > 0$. Assume that $\A$ satisfies robust injectivity on $\Gamma K_s^{5R}$ with injectivity constants $\gamma \in (0,1]$ and $\delta > 0$. If $\X_\star \in \Gamma K_{s}^{R}$ and $\y = \A(\X_\star) + \Eta~\in~R^m$, then by choosing $\Alpha, \Beta$ sufficiently small with $\alpha_1 = \sqrt{\frac{s}{\Gamma}} \alpha_2 = \beta_1 = \sqrt{\frac{s}{\Gamma}} \beta_2$ one obtains
    \begin{align} \label{Approximation}
        \Vert \X_\star - \X_\abg \Vert_F \le \frac{2}{\sqrt{\gamma}} \norm{\Eta}{2} + \sqrt{\delta},
    \end{align}
    for any global minimizer $(\U_\abg, \V_\abg)$ of $J_\abg^R$. In particular, $\X_\abg \in \Gamma K_{s}^{4R}$ with the SD in \eqref{Xab}.

    \rev{If $\X_\star \in \bar{S}_s^R$ and $\alpha_1 = \sqrt{\frac{s}{\Gamma}} \alpha_2 = \beta_1 = \sqrt{\frac{s}{\Gamma}} \beta_2$, it suffices that $\A$ satisfies robust injectivity on $\tfrac{\| \X_\star \|_F}{R} K_s^{5R}$ to obtain \eqref{Approximation}. In this case, $\X_\abg \in \tfrac{\| \X_\star \|_F}{R} K_{s}^{4R}$ with the SD in \eqref{Xab}.}
\end{theorem}
\begin{Proof}
    By assumption $\X_\star \in \Gamma K_s^R$ and $\alpha_1 = \sqrt{\frac{s}{\Gamma}} \alpha_2 = \beta_1 = \sqrt{\frac{s}{\Gamma}} \beta_2$. According to Lemma \ref{Bound-y2} we can choose $(\Alpha,\Beta)$ such that $\norm{\Eta}{2} \le \norm{\y - \A(\X_\abg)}{2} \le 2\norm{\Eta}{2}$. Lemma \ref{Bound-uv2} thus yields that $\X_\abg \in \Gamma K_s^{4R}$. Since $\X_\star - \X_\abg \in \Gamma K_s^{5R}$ by Lemma \ref{lemma:sumrule}, we can now apply the robust injectivity of $\A$ to obtain
	\begin{align*}
		\| \X_\star - \X_\abg \|_{F} 
		&\le \frac{1}{\sqrt{\gamma}} \norm{\y - \A(\X_\abg)}{2} + \sqrt{\delta} \\
		&\le \frac{2}{\sqrt{\gamma}} \norm{\Eta}{2} + \sqrt{\delta}.
	\end{align*}
	and hence the first claim.

    \rev{Assume now that $\X_\star \in \bar{S}_s^R$. As described in the context of \eqref{SVD}, $\X_\star$ hence possesses an orthogonal SD $\X_\star = \U_\star \V_\star^\top$ with $\| \U_\star \|_F = \| \V_\star \|_F = \| \X_\star \|_F^{\frac{1}{2}}$ and $\| \U_\star \|_1, \| \V_\star \|_1 \le \| \X_\star \|_F^{\frac{1}{2}} \sqrt{sR}$ (the $\ell_1$-norm bounds follow from $\U_\star, \V_\star$ having at most $Rs$ entries). Recalling \eqref{eq:K}, it clearly follows from 
    \begin{align*}
        \X_\star = \frac{\| \X_\star \|_F}{R} \Big( \frac{\sqrt{R}}{\| \X_\star \|_F^\frac{1}{2}} \U_\star \Big) \Big( \frac{\sqrt{R}}{\| \X_\star \|_F^\frac{1}{2}} \V_\star \Big)^\top = \frac{\| \X_\star \|_F}{R} \tilde{\U}_\star \tilde{\V}_\star^\top
    \end{align*}
    that $\X_\star \in \tfrac{\| \X_\star \|_F}{R} K_s^{R}$. This obviously yields the second claim by using the same arguments as before.}
\end{Proof}
\rev{Whereas the assumptions of Theorem \ref{ApproxX} are standard, i.e., the measurements need to behave well on the signal set and the parameters need to be well-tuned depending on the signal norm and noise level, the error bound in \eqref{Approximation} is less common. Indeed, the additive appearance of $\sqrt{\delta}$ means that the Frobenius ball $B$ of radius $\sqrt{\delta}$ centered at the origin is a ``dead zone'': For any signal $\X_\star \in B$, the statement of Theorem \ref{ApproxX} is void. As long as all signals of interest are of a similar order of magnitude though, $\delta$ can be considered to be small in comparison by assuming a respectively strong injectivity property of $\A$. The results in Section \ref{SectionRIP} illustrate how this relates to a sufficient oversampling factor in the number of observations. If applied in a non-uniform way, Theorem \ref{ApproxX} thus yields reconstruction of single signals of exceedingly small magnitude.  

A second implication of the non-standard error bound in \eqref{Approximation} is that the reconstruction error is not converging to zero for vanishing noise. Although this seems counter-intuitive, the experiments in Section \ref{Numerics3} suggest that the additive term in \eqref{Approximation} might be factual for general ground-truths in $K_{s_1,s_2}^R$. In the case of jointly sparse ground-truths in $\bar{S}_{s_1,s_2}^R$, the experiments clearly show that it is an artifact of the proof since the reconstruction error vanishes with the noise.}

The main challenge in applying Theorem \ref{ApproxX} is to tune the parameters $\Alpha$ and $\Beta$. This is common to various regularized recovery procedures. Lemma \ref{Bound-uv2}, nevertheless, suggests a simple heuristic. As long as minimizers of $J_\abg^R$ are regular in the sense that the (effective) sparsity of $\U_\abg$ and $\V_\abg$ is small, the critical point of $\norm{\y - \A(\X_\abg)}{2} = \norm{\Eta}{2}$ has not yet been reached and $\Alpha, \Beta$ may be further decreased. If the critical point $\norm{\y - \A(\X_\abg)}{2} < \norm{\Eta}{2}$ is hit, the (effective) sparsity is not controlled anymore and likely to increase massively. This behavior can be observed in numerical experiments, see Section \ref{Numerics}.

\rev{Finally, when considering subgaussian measurements, the second part of Theorem \ref{ApproxX} allows us to reach with our method the same sample complexity as \cite{lee2013near}, cf.\ \eqref{eq:SubgaussRemark} in Remark \ref{rem:dim}.}


\subsection{Robust Injectivity} \label{SectionRIP}

\rev{
To make use of Theorem \ref{ApproxX}, we need to understand under which assumptions a linear operator $\A$ of the form \eqref{A} satisfies the robust injectivity property introduced in Definition \ref{apprRIPDef}. If the component matrices $\AA_i$ are constructed as i.i.d.\ copies of a random matrix $\AA$, it suffices that the probability mass is not strongly concentrated around zero. This may be formally expressed in the \emph{small ball} estimate
\begin{align} \label{eq:SmallBall}
	\Pr{\inner{\AA,\Z}{F}^2 \ge \theta \norm{\Z}{F}^2 } \ge c, \quad \text{ for all } \Z \in K,
\end{align}
where $K \subset \R^{n_1\times n_2}$ is the signal set of interest and $\theta,c > 0$ are absolute constants. Note that \eqref{eq:SmallBall} does not require independent entries of $\AA$. It is straight-forward to deduce for such $\A$ the following result from Mendelson's work in \cite{mendelson2014learning}.
\begin{theorem}[Robust injectivity for small ball operators] \label{thm:SmallBallRIP}
	Let $\Gamma > 0$, \rev{$\delta > 0$,} and $\A \colon \R^{n_1\times n_2} \rightarrow \R^m$ be a linear measurement operator of the form \eqref{A} whose components $\AA_i$ are i.i.d.\ copies of a random matrix $\AA$ satisfying \eqref{eq:SmallBall} on $\Gamma K_{s_1,s_2}^R - \Gamma K_{s_1,s_2}^R$. If
	\begin{align} \label{eq:measSmallBall}
	    m \gtrsim \delta^{-2} E_m (\Gamma K_{s_1,s_2}^R - \Gamma K_{s_1,s_2}^R, \AA)^2,
	\end{align}
	where 
	\begin{align} \label{eq:EmpiricalWidth}
		E_m (K,\AA) = \E{ \sup_{\Z \in K} \inner{\frac{1}{\sqrt{m}}  \sum_{i = 1}^m \eps_i \AA_i, \Z }{F} }
	\end{align}
	denotes the mean empirical width of a set $K \subset \R^{n_1 \times n_2}$ under the distribution of $\AA$ (the $\eps_i$ are i.i.d.\ copies of a Rademacher variable), then $\A$ satisfies with probability at least $1-\exp(-\frac{c^2}{2} m)$ robust injectivity as in \eqref{apprRIP} with constants $\gamma = \frac{\theta^2 c}{16}$ and $\delta$.
\end{theorem}
The proof of Theorem \ref{thm:SmallBallRIP} can be found in Section \ref{sec:SmallBallRIPProof}. The result can be applied to various choices of $\A$, e.g., measurement operators characterized by subgaussian matrices $\AA_i$ exhibiting strong tail decay, measurement operators whose components $\AA_i$ may be heavy tailed but obey an $L_1$-$L_2$-equivalence, or measurement operators whose components are Gaussian rank-1 matrices, i.e., $\AA_i = \a_i\a_i^\top$ where $\a_i \in \R^n$ has i.i.d.\ Gaussian entries. In the following sections we will discuss the implications of Theorem \ref{thm:SmallBallRIP} in each of these settings.

\subsubsection{Subgaussian measurement operator with isotropic components}
\label{sec:SubgaussianDetails}
We first recall the definition of subgaussian random variables and vectors. 

\begin{definition}[Subgaussian random variable]
\label{def:subgaussian}
	A random variable $\xi \in \R$ is called $\mathcal{K}$-subgaussian if the tail bound $\Pr[]{|\xi| > t} \le 2\exp(-t^2/\mathcal{K}^2)$ holds. The smallest possible number for $\mathcal{K} \ge 0$ is called subgaussian norm of $\xi$ and denoted by $\Vert \xi \Vert_{\psi_2}$. 
\end{definition}
The class of subgaussian random variables covers important special cases such as Gaussian, Bernoulli, and more generally all bounded random variables, cf.\ \cite{vershynin2010introduction}. Furthermore, any subgaussian random variable $\xi \in \R$ satisfies
\begin{align} \label{eq:MomentCondition}
    \| \xi \|_{L_p} \lesssim \mathcal K \sqrt{p},
    \qquad \text{for all } p \ge 1,
\end{align}
where $\| \xi \|_{L_p} := \E{|\xi|^p}{}^{\frac{1}{p}}$ denotes the $p$-th absolute moment of $\xi$. The condition in \eqref{eq:MomentCondition} is (up to scaling $\mathcal K$ by an absolute constant) equivalent to the tail decay in Definition \ref{def:subgaussian}. Subgaussian random vectors can now be defined as follows.

\begin{definition}[Subgaussian random vector]
    A random vector $\boldsymbol \xi \in \R^n$ is called $\mathcal K$-subgaussian if the random variable $\langle \x, \boldsymbol \xi \rangle$ is $\mathcal K$-subgaussian, for any choice of $\x \in \R^n$.
\end{definition}

A random vector $\boldsymbol \xi \in \R^n$ is called isotropic if $\| \langle \x, \boldsymbol \xi \rangle \|_{L_2} = \| \x \|_2$, for any $\x \in \R^n$. Since this is equivalent to $\EE[\boldsymbol\xi \boldsymbol\xi^\top] = \id$, isotropic random vectors are a natural generalization of unit variance random variables to higher dimensions. Both concepts (subgaussianess and isotropy) extend to random matrices by interpreting them as vectors.

For subgaussian measurement operators, i.e., operators $\A$ whose component matrices $\AA_i$ are i.i.d.\ copies of a subgaussian matrix $\AA$, the mean empirical width in \eqref{eq:EmpiricalWidth} reduces to the so-called \emph{Gaussian width} defined as
\begin{align*}
	w(K) = \E{\sup_{\z \in K} \inner{\z,\g}{} },
\end{align*}
for $K \subset \R^n$ and $\g \in \R^n$ having i.i.d.\ standard Gaussian entries, cf.\ \cite[Eq.\ (6.4)]{tropp2015convex}.
The Gaussian width has become an established measure of set complexity due to its numerous favorable properties \cite{vershynin2018high}. In particular, 
\begin{itemize}
    \item the width of the Euclidean unit ball $B^{n}$ fulfills $w(B^n) \simeq \sqrt{n}$,
    \item the width of the set of $s$-sparse vectors $\Sigma_s^n$ intersected with $B^n$ fulfills $w(\Sigma_s^n \cap B^n) \simeq \sqrt{s \log\round{\frac{en}{s}}}$,
\end{itemize} 
illustrating that $w(K \cap B^n)^2$ extends the linear dimension of subspaces in a consistent way to arbitrary sets $K \subset \R^n$. The restriction to $B^n$ in the above examples is necessary since $w(K)$ scales in the diameter of $K$. By noticing that isotropic subgaussian random vectors satisfy \eqref{eq:SmallBall} and by evaluating $w(\Gamma K_{s_1,s_2}^R - \Gamma K_{s_1,s_2}^R)$, we can deduce the following corollary of Theorem \ref{thm:SmallBallRIP}. The details are provided in Section \ref{sec:coveringnumbers}.
\begin{corollary}[Robust injectivity for subgaussian operators] \label{GaussianRIP}
	Let $\Gamma > 0$, \rev{$\delta > 0$,} and $\A \colon \R^{n_1\times n_2} \rightarrow \R^m$ be a linear measurement operator of form \eqref{A}. Assume, all $\AA_i$, for $i \in [m]$, are i.i.d.\ copies of a $\mathcal{K}$-subgaussian isotropic random matrix $\AA$. If
	\begin{align} \label{meas1}
	     m \gtrsim \delta^{-2} \Gamma^2 R^3 (s_1 + s_2) \log^3 \round{ \max \curly{n_1, n_2} },
	\end{align}
	then $\A$ satisfies with probability at least $1-\exp(-\frac{c^2}{2} m)$ robust injectivity as in \eqref{apprRIP} with constants $\gamma$ and $\delta$, where $\gamma$ solely depends on $\mathcal K$.
    
\end{corollary}

}

%
\begin{remark} \label{DominiksRemark}
	\rev{
	As observed in \cite{fornasier2018robust}, the strong concentration of subgaussian random variables around their mean even allows to obtain the stronger bound
	\begin{align} \label{apprRIP2}
	    \left| \Vert \A(\Z) \Vert_2^2 - \Vert \Z \Vert_F^2 \right| \le \delta,
	\end{align}
	for any $\Z \in \Gamma K_{s_1,s_2}^R$, which implies robust injectivity with $\gamma = 1$ and $\delta$ (note at this point, however, that $\Gamma K_{s_1,s_2}^R$ differs from the signal set in \cite{fornasier2018robust} and that Corollary \ref{GaussianRIP} shows a better sample complexity than \cite[Lemma 5.3]{fornasier2018robust}).
	The additive quasi-isometric property in \eqref{apprRIP2} is still not of the form of commonly used multiplicative \emph{Restricted Isometry Properties (RIP)}
	\begin{align} \label{multRIP}
	    (1-\delta) \Vert \Z \Vert_F^2 \le \Vert \A(\Z) \Vert_2^2 \le (1+\delta) \Vert \Z \Vert_F^2
	\end{align}
	as \eqref{apprRIP2}, in contrast to \eqref{multRIP}, is not scaling invariant and $\A(\Z) = \A(\Z')$ does not imply $\Z = \Z'$ but only $\Vert \Z - \Z' \Vert_2^2 \le \delta$.
	
	Let us repeat in this context an important observation from \cite{fornasier2018robust}. In fact, it is \emph{impossible} to derive a classical scaling invariant RIP like \eqref{multRIP} on $\Gamma K_{s_1,s_2}^{R}$ under similar conditions as \eqref{meas1}. The main problem is non-orthogonality of the SDs in $\Gamma K_{s_1,s_2}^{R}$. This can be seen in the following way (we consider  w.l.o.g.\ only the case $R=2$): Let $m \simeq 2 (n_1+s) \log^3 \left( \max \{n_1,n_2\} \right)$ and assume that the linear operator $\A$ fulfills \eqref{multRIP} for all $\Z \in K_{n_1,s}^{2}$. Let $\u \in \R^{n_1}$ and $\w \in \R^{n_2}$ be arbitrary unit-norm vectors. Now choose any $\v_1 \in \R^{n_2}$ of unit norm and $\Vert \v_1 \Vert_1 \le \sqrt{s}/2$. Define $\v_2 := -\v_1 + \eps \w$ and choose $\eps > 0$ sufficiently small to ensure $\Vert \v_2 \Vert_1 \le \sqrt{s}$ and $\Vert \v_2 \Vert_2 \approx 1$. Then $\Z := (1/2)\u\v_1^\top + (1/2)\u\v_2^\top \in K_{n_1,s}^{2}$ and, by assumption, \eqref{multRIP} holds. But this implies by definition of $\Z$ and scaling invariance of \eqref{multRIP} that
	\begin{align} \label{eq:LowRankRIP}
		(1-\delta) \Vert \u\w^\top \Vert_F^2 \le \Vert \A(\u\w^\top) \Vert_2^2 \le (1+\delta) \Vert \u\w^\top \Vert_F^2.
	\end{align}
	Since the choice of $\u$ and $\w$ was arbitrary, this means that \eqref{eq:LowRankRIP} and thus also \eqref{multRIP} hold for all rank-$1$ matrices (not only those with effectively sparse right components). If $n_1,s \ll n_2$, this is a clear contradiction to information theoretical lower bounds, as corresponding RIPs would require at least $m \simeq \max\{n_1,n_2\}$ (see \cite[Section 2.1]{candes2011tight}).

    Finally note that even if one used the alternative and scaling-invariant definition of effective sparsity mentioned in Remark \ref{rem:EffectiveSparsity} to define the set $K_{s_1,s_2}^R$, the above argument would stay valid since $\v_1$ and $\v_2$ are of unit norm and thus effectively $s$-sparse according to both definitions. The main obstacle in obtaining scaling invariant RIPs like \eqref{multRIP} is thus the non-orthogonality of decompositions in $K_{s_1,s_2}^R$, \emph{not} the scaling variance of our effective sparsity model.
	}
\end{remark} 
\rev{
\begin{remark}\label{rem:dim}
	Note that $\Gamma R$ is the Frobenius diameter of $\Gamma K_{s_1,s_2}^{R}$ and that \eqref{meas1} can be re-written as
	\begin{align} \label{meas1_alternativeForm}
	     m \gtrsim \round{\frac{\delta}{\Gamma R}}^{-2} R (s_1 + s_2) \log^3 \round{ \max \curly{n_1, n_2} }.
	\end{align}
	For $\delta = \Delta \Gamma R$, $\Delta \in (0,1)$, Corollary \ref{GaussianRIP} hence states that up to log-factors $m \approx \mathcal{O} \left( \Delta^{-2} R(s_1+s_2)) \right)$ subgaussian measurements are sufficient to $\delta$-stably embed $\Gamma K_{s_1,s_2}^{R}$ in $\R^m$, cf.\ \cite[Def. 1.1 \& Thm. 1.5]{Plan2014}. 
	How does this relate to the preliminary work on SPF in \cite{lee2013near} where orthogonality of the matrices $\U$ and $\V$ was assumed? First note that to allow a fair comparison between the scaling invariant result in \cite{lee2013near} and the scaling variant Corollary \ref{GaussianRIP} (cf.\ Remark \ref{DominiksRemark}), one needs to restrict oneself to unit Frobenius-norm matrices $\norm{\Z}{F} = 1$. The results in \cite{lee2013near} now guarantee that \eqref{multRIP} holds for all $\Z \in \bar{S}_{s_1,s_2}^R$ if $m \gtrsim \delta^{-2} R (s_1 + s_2) \log(\max\{n_1,n_2\})$. Let us compare this to our results. If $\Z \in \bar{S}_{s_1,s_2}^R$ is of unit Frobenius-norm, the proof of Theorem \ref{ApproxX} shows that $\Z \in \frac{1}{R} K_{s_1,s_2}^R$. \rev{Applying Corollary \ref{GaussianRIP} to this set, i.e., setting $\Gamma = \tfrac{1}{R}$, yields that the robust injectivity holds with high probability if}
	\begin{align} \label{eq:SubgaussRemark}
		m \gtrsim \delta^{-2} R (s_1 + s_2) \log^3 \round{ \max \curly{n_1, n_2} }.
	\end{align} 
	Up to the additional log-factors, this perfectly resembles the results in \cite{lee2013near}, which have been shown to be near-optimal for recovery of matrices with orthogonal sparse decomposition. However, \cite{lee2013near} neither treats effective sparsity nor non-orthogonal decompositions.
\end{remark}
}
\subsubsection{Heavy-tailed measurements with $L_1$-$L_2$-equivalence} 
\label{sec:L1L2equivalence}

\rev{Theorem \ref{thm:SmallBallRIP} does not require strong concentration properties as fulfilled by subgaussian random matrices. It doesn't even require the existence of higher-order moments. We can, for instance, consider operators $\A$ of form \eqref{A} whose component matrices $\AA_i$ are i.i.d.\ copies of an isotropic random matrix $\AA$ that only satisfies the $L_1$-$L_2$-equivalence}
\begin{align} \label{eq:L1L2}
	\E{\inner{\AA,\Z}{F}^2}^{\frac{1}{2}} \le c \E{\abs{\inner{\AA,\Z}{F}}{}}, \quad \text{ for all } \Z \in \R^{n_1 \times n_2},
\end{align}
where $c > 1$ is an absolute constant. The condition in \eqref{eq:L1L2} is fulfilled by various heavy-tailed distributions. By isotropy and the Paley-Zygmund inequality \cite[Lemma 4.1]{mendelson2014learning}, one obtains from \eqref{eq:L1L2} the small ball estimate
\begin{align} \label{eq:SmallBallHeavy}
	\Pr{\inner{\AA,\Z}{F}^2 \ge \theta \norm{\Z}{F}^2 } \ge \round{\frac{1 - \sqrt{\theta} c}{c}}^2, \quad \text{ for all } \Z \in \R^{n_1 \times n_2},
\end{align}
for all $\theta \in \round{ 0,\frac{1}{c^2} }$, \rev{and Theorem \ref{thm:SmallBallRIP} can be applied. Note, however, that the required sample complexity depends on $E_m (\Gamma K_{s_1,s_2}^R - \Gamma K_{s_1,s_2}^R, \AA)$, which is hard to compute for general $\AA$. The experiments in Section \ref{sec:Numerics2} suggest that the sample complexity for log-normal $\AA$ is comparable to the Gaussian case.}
%
%
\begin{remark}
	By \cite[Lemma 4.2]{mendelson2014learning} the estimate in \eqref{eq:SmallBall} can be deduced as well if $\AA$ consists of i.i.d.\ entries $A_{i,j}$ which satisfy $\E{A_{i,j}^2} \le c \E{\abs{A_{i,j}}{}}$ (only the constant on the right-hand side of \eqref{eq:SmallBall} changes). However, the condition in \eqref{eq:L1L2} is more general as it does not require independence of the entries of $\AA$.
\end{remark}

\subsubsection{Rank-1 measurement operator} 
Another noteworthy class of measurement operators for matrix sensing is given by rank-$1$ measurements, i.e., the matrices $\AA_i$ defining $\A$ are i.i.d.\ copies of a rank one matrix $\AA = \a\a^\top$ where $\a \in \R^n$ has i.i.d.\ standard Gaussian entries (in this case we restrict ourselves to quadratic matrices $n_1 = n_2 = n$). The advantage of such a structured measurement operator lies in the reduced storage and evaluation costs. The drawback when compared to subgaussian operators with i.i.d.\ entries is the additional dependence between the entries of each $\AA_i$. However, by using the observation that Gaussian rank-$1$ measurements satisfy the small ball estimate \eqref{eq:SmallBall}, cf.\ \cite{kueng2017low}, one may again apply Theorem \ref{thm:SmallBallRIP}. As in Section \ref{sec:L1L2equivalence}, computing $E_m (\Gamma K_{s_1,s_2}^R - \Gamma K_{s_1,s_2}^R, \AA)$ is non-trivial but the numerical experiments in Section \ref{sec:Numerics2} suggest that it scales comparable to the subgaussian case. 

The following lemma, which relates $E_m (\Gamma K_{s_1,s_2}^R - \Gamma K_{s_1,s_2}^R, \AA)$ to Talagrand's $\gamma_1$- and $\gamma_2$-functionals, can be viewed as a first step in deriving an explicit sample complexity. The $\gamma_\alpha$-functional for a metric space $(K,\mathrm{dist})$ is defined as
\begin{align} \label{eq:gamma_2}
	\gamma_\alpha(K,\mathrm{dist}) = \inf_{(K_i)_{i \in \mathbb{N}}} \sup_{\z \in K} \sum_{i = 0}^\infty 2^{\frac{i}{\alpha}} \mathrm{dist}(\z,K_i),
\end{align}
where the infimum is taken over all \emph{admissible sequences} $(K_i)_{i \in \mathbb{N}}$, i.e., sequences of subsets of $K$ for which $\abs{K_0}{} = 1$ and $\abs{K_i}{} \le 2^{2^i}$.

\begin{lemma}
\label{lem:Talagrand}
    For $K \subset \R^{n\times n}$ and $\AA = \a\a^\top$ with $\a \in \R^n$ having i.i.d.\ $\mathcal{K}$-subgaussian entries with mean $0$ and variance $1$, one has
    \begin{align}
    \label{eq:Talagrand}
        E_m(K,\AA) \le C \round{ \sqrt{R} \gamma_2(K,\norm{\cdot}{F}) + \frac{\gamma_1(K,\norm{\cdot}{})}{\sqrt{m}} },
    \end{align}
    where $C > 0$ only depends on $\mathcal{K}$.
\end{lemma}
%

The proof of Lemma \ref{lem:Talagrand} can be found in Section \ref{sec:Rank1measProof}. Note that $\gamma_\alpha(K,\mathrm{dist})$ can be seen as a measure of the intrinsic complexity of $K$ when measured in $\mathrm{dist}$. Up to an absolute constant $\gamma_2(K,\norm{\cdot}{2})$ is equivalent to the Gaussian width \cite[Theorem 8.6.1]{vershynin2018high}, a result due to Fernique \cite{fernique1975regularite} and Talagrand \cite{talagrand2006generic}, and widely known as \emph{majorizing measures} theorem.
By this result, one can replace $\gamma_2(\Gamma K_{s_1,s_2}^R - \Gamma K_{s_1,s_2}^R,\norm{\cdot}{F})^2$ in Lemma \ref{lem:Talagrand} with $w(\Gamma K_{s_1,s_2}^R - \Gamma K_{s_1,s_2}^R)^2$ leading to a bound as in \eqref{meas1}. Unfortunately, it is not as simple to bound $\gamma_1(\Gamma K_{s_1,s_2}^R - \Gamma K_{s_1,s_2}^R,\norm{\cdot}{})^2$ in an intuitive way without deriving a tight bound on the covering number $N(\Gamma K_{s_1,s_2}^R,\| \cdot \|, \varepsilon)$ of $\Gamma K_{s_1,s_2}^R$ in operator norm. \rev{Indeed, if we had such a bound, we could use that
\begin{align*}
    \gamma_1(\Gamma K_{s_1,s_2}^R - \Gamma K_{s_1,s_2}^R,\norm{\cdot}{})
    \lesssim \int_0^\infty \log\big( N(K_{s_1,s_2}^R - \Gamma K_{s_1,s_2}^R,\| \cdot \|, \varepsilon) \big) d\varepsilon,
\end{align*}
see \cite[Section 1.2]{talagrand2006generic}, to control the second term in \eqref{eq:Talagrand}.} Independently from this open point, let us mention that in its current form we do not expect \eqref{eq:Talagrand} to be optimal since the additional factor $\sqrt{R}$ in front of the $\gamma_2$-functional appears to be an artifact of the proof.


\subsection{Computing Minimizers}
\label{sec:Convergence}

The previous sections showed that minimizers of $J_\abg^R$ uniformly approximate original ground-truths as long as the measurement operator $\A$ satisfies the robust injectivity in Definition \ref{apprRIPDef}, a rather mild condition fulfilled by many popular choices of $\A$. The last crucial question is how to compute global minimizers of the functional $J_\abg^R$ defined in \eqref{Jab0} in an efficient way. We discuss here two schemes based on the proximal operator of the elastic net $\EN_\THeta(\Z) = \theta_1 \norm{\Z}{F}^2 + \theta_2 \norm{\Z}{1}$, for $\THeta = (\theta_1,\theta_2)^\top \ge \0$. These methods are designed to handle the non-differentiability of $J_\abg^R$ which is inherited from the $\ell_1$-norm. The proximal operator of a proper and lower semi-continuous function $f \colon \R^n \rightarrow \R \cup \curly{\infty}$ is defined as
\begin{align*}
	\prox{\mu f}(\u) = \argmin_{\z \in \R^n} \; \frac{1}{2} \norm{\z - \u}{2}^2 + \mu f(\z),
\end{align*}
where $\mu > 0$ is a design parameter. Since $\prox{\chi_C}(\z)$ is the orthogonal projection of $\z$ onto $C$, for $C$ convex and $\chi_C$ being the corresponding indicator function \cite{rockafellar2009variational}, proximal operators may be viewed as generalized projections. By separability of the components of $\EN_\THeta$ it is straight-forward to verify that
\begin{align*} 
	\prox{\mu \EN_\THeta} (\Z) = \round{
		S_{\THeta, \mu} (Z_{i,j}) }_{i,j} 
	\tab 
	\text{ where } S_{\THeta, \mu} (Z_{i,j}) = \begin{cases}
		\frac{Z_{i,j} - \mu \theta_2}{1 + 2\mu  \theta_1} & Z_{i,j} > \mu \theta_2 \\
		0 & |Z_{i,j}| \le \mu \theta_2 \\
		\frac{Z_{i,j} + \mu \theta_2}{1 + 2\mu \theta_1} & Z_{i,j} < -\mu \theta_2
	\end{cases}.
\end{align*}
Due to the non-convex structure of $J_\abg^R$, we are not able to derive guaranteed global convergence to global minimizers. However, we guarantee global convergence of all presented methods to stationary points of $J_\abg^R$ and local convergence to global minimizers. Our analysis relies on the fact that $J_\abg^R$ has the so called Kurdyka-Lojasiewicz property, which requires $J_\abg^R$ to behave well around stationary points. 
\begin{definition}[{Kurdyka-Lojasiewicz property \rev{\cite[Definition 3.1]{attouch2010proximal}}}]
	A proper lower semicontinuous function $f:\mathbb{R}^n \rightarrow {\mathbb{R} \cup \{ \infty \}}$
	is said to have the KL-property at $\ol{\x} \in \dom \, \partial f$ if there exist $\eta \in 
	\left( 0,\infty \right]$, a neighborhood $U$ of $\overline{\x}$ and a continuous concave
	function $\varphi : \left[ 0,\infty \right) \rightarrow \mathbb{R}_+$ such that
	\begin{enumerate}
		\item[-] $\varphi(0) = 0$,
		\item[-] $\varphi$ is $C^1$ on $(0,\eta)$, 
		\item[-] $\varphi'(t) > 0$, for all $t \in (0,\eta)$,
		\item[-] and, for all $\x \in U \cap \{ \x \in \mathbb{R}^n : f(\overline{\x}) < f(\x) <
		f(\overline{\x}) + \eta \}$, the KL-inequality holds:
		\begin{align*}
		\varphi'(f(\x)-f(\overline{\x}))\; \dist(0,\partial f(\x)) \geq 1.
		\end{align*}
	\end{enumerate}
\end{definition}
\begin{remark}{} \label{rem:Radius}
    \rev{As mentioned in \cite{attouch2010proximal}, the KL-property originates from \cite{lojasiewicz1963propriete}, where the KL-inequality was proven for real-analytic functions and $\varphi(t) = t^{1-\theta}$, for $\theta \in [\tfrac{1}{2},1)$. Consequent works \cite{kurdyka1998gradients,bolte2007lojasiewicz} extended the result to non-smooth subanalytic functions. It is easy to see that $J_\abg^R$ has the KL-property with $\varphi(t) = ct^{1-\theta}$, for $c > 0$ and $\theta \in [0,1)$ since its graph is a semialgebraic set in $\R^{n_1 \times R} \times \R^{n_2 \times R} \times \R$, see the more detailed discussion in Section \ref{sec:ConvergenceProof}.} 
    
    As \rev{\cite[Theorem 3.4]{attouch2010proximal}} shows, a characterization of $\theta$ would determine the convergence speed of alternating descent schemes. While \cite{li2013global} can be used to compute $\theta$ for piecewise convex polynomials, it is unclear how to do the same for non-convex polynomials. Addressing this more general issue would, in particular, provide convergence rates for \eqref{ATLAS0}.
    
	Similarly, the main difficulty in characterizing the convergence radius in Theorems \ref{LocalConvergence_AM} and \ref{LocalConvergence_PALM} below is to characterize the KL-parameters $U$ and $\eta$ of $J_\abg^R$. Doing so for a non-convex functional is a challenging task on its own and thus the reason for us to defer the treatment of initialization to future work.
\end{remark}

\subsubsection{Alternating Minimization} 

%

\rev{To analyze \eqref{ATLAS0}, i.e., Algorithm \ref{alg:AM}, we follow the arguments in \cite{attouch2010proximal}. Note, however, that Algorithm~\ref{alg:AM} is rather inefficient since each iteration requires solving a convex optimization problem. If desired, this could be done, e.g., via proximal gradient descent given in Algorithm~\ref{alg:ProxDesc}.}

\begin{theorem} \label{LocalConvergence_AM}
	The sequence $(\U_k,\V_k)$ generated by Algorithm \ref{alg:AM} converges to a stationary point of $J_\abg^R$.
	Moreover, for any global minimizer $(\U_\abg,\V_\abg)$ of $J_\abg^R$, there exist $\eps, \eta > 0$, such that the initial conditions
	\begin{align*}
		\Vert (\U_0,\V_0) - (\U_\abg,\V_\abg) \Vert_F < \eps, \tab \min J_\abg^R < J_\abg^R(\U_0,\V_0) < \min J_\abg^R + \eta,
	\end{align*}
	imply that the iterations $(\U_k,\V_k)$ converge to some $(\U_*,\V_*) \in \argmin J_\abg^R$.
\end{theorem}
\rev{We discuss in Section \ref{sec:ConvergenceProof} how Theorem \ref{LocalConvergence_AM} can be derived from the results in \cite{attouch2010proximal}.}
\begin{algorithm} 
	\caption{\textbf{:}  \textbf{Alternating Minimization}} \label{alg:AM}
	\begin{algorithmic}[1]
		\Require{$\y \in \R^m$, $\AA \in \R^{m\times n_1n_2}$ such that $\A(\U\V^\top) = \AA \cdot \vect(\U\V^\top)$, rank $R$, $\V_0 \in \R^{n_2 \times R}$, and $\alpha_1,\alpha_2,\beta_1,\beta_2 > 0$}
		\Statex
		\While{stop condition is not satisfied}
		\Let{$\U_{k+1}$}{$\argmin_{\U \in \R^{n_1 \times R}} J_\abg^R(\U,\V_k)$}
		\Comment{Use, e.g., Algorithm \ref{alg:ProxDesc}}
		\Let{$\V_{k+1}$}{$\argmin_{\V \in \R^{n_2 \times R}} J_\abg^R(\U_{k+1},\V)$}
		\Comment{Use, e.g., Algorithm \ref{alg:ProxDesc}}
		\EndWhile
		\State 
		\Return{$(\U_\text{final},\V_\text{final})$}
	\end{algorithmic}
\end{algorithm}
\begin{algorithm} 
	\caption{\textbf{:}  \textbf{Proximal Gradient Descent}} \label{alg:ProxDesc}
	\begin{algorithmic}[1]
		\Require{objective function $h = f + g \colon \R^n \rightarrow \R$ with $f$ continuously differentiable and $g$ lower semicontinuous and convex, $\x_0 = \0$, step-size $\mu > 0$}
		\Statex
		\While{stop condition is not satisfied}
		\Let{$\x_{k+1}$}{$\prox{g}(\x_k - \mu \nabla f(\x_k))$}
		\EndWhile
		\State 
		\Return{$\x_\text{final}$}
	\end{algorithmic}
\end{algorithm}

\subsubsection{Proximal Alternating Linearized Minimization}

A second, more efficient descent scheme is given by the Proximal Alternating Linearized Minimization \cite{bolte2014proximal}, see Algorithm \ref{alg:PALM}. \rev{The matrices $\AA_{\V_k} \in \R^{m \times Rn_1}$ and $\AA_{\U_{k+1}} \in \R^{m \times Rn_2}$ appearing in Algorithm \ref{alg:PALM} are defined via 
\begin{align*}
    \A(\U\V_k^\top) = \AA_{\V_k} \cdot \vect(\U)
    \quad \text{and} \quad 
    \A(\U_{k+1} \V^\top) = \AA_{\U_{k+1}} \cdot \vect(\V),
\end{align*}
for any $\U \in \R^{n_1 \times R}$ and $\V \in \R^{n_2 \times R}$.
We verify in Section \ref{sec:ConvergenceProof} that the coercive functional $J_\abg^R$ and Algorithm~\ref{alg:PALM} satisfy Assumptions 1\&2 in \cite{bolte2014proximal}. Consequently, the following holds.}
\begin{theorem}[{\cite[Lemma 3, Theorem 1]{bolte2014proximal}+\rev{\cite[Theorem 3.3]{attouch2010proximal}}}] \label{LocalConvergence_PALM}
	The sequence $(\U_k,\V_k)$ generated by Algorithm~\ref{alg:PALM} converges to a stationary point of $J_\abg^R$.
	Moreover, for any global minimizer $(\U_\abg,\V_\abg)$ of $J_\abg^R$, there exist $\eps, \eta > 0$, such that the initial conditions
	\begin{align*}
		\Vert (\U_0,\V_0) - (\U_\abg,\V_\abg) \Vert_F < \eps, \tab \min J_\abg^R < J_\abg^R(\U_0,\V_0) < \min J_\abg^R + \eta,
	\end{align*}
	imply that the iterations $(\U_k,\V_k)$ converge to some $(\U_*,\V_*) \in \argmin J_\abg^R$.
\end{theorem}
\begin{remark}
\label{rem:Haeffele}
    \rev{When one tries to combine Theorems \ref{LocalConvergence_AM} and \ref{LocalConvergence_PALM} with Theorem \ref{ApproxX} to obtain practical reconstruction guarantees, one encounters the main weakness of these results. The parameters $\varepsilon$ and $\eta$, which determine the local convergence radius of Alorithms \ref{alg:AM} and \ref{alg:PALM}, depend on the KL-property of $J_\abg^R$ and are thus hard to determine, cf.\ Remark \ref{rem:Radius}. To obtain meaningful statements in light of the ``dead zone'' of Theorem \ref{ApproxX}, (see the attached discussion after the theorem) one however requires $\varepsilon > \sqrt{\delta}$. An interpretable characterization of the convergence radius is thus highly desirable.
    
    An alternative approach to obtain such practical recovery guarantees might be to use instead of Algorithm~\ref{alg:AM} or \ref{alg:PALM} the meta-algorithm \cite[Algorithm 1]{haeffele2019structured} which provably converges to global minimizers of functionals like $J_\abg^R$ if one has a descent procedure at hand that provably converges to local minima of the same. Unfortunately, Algorithms \ref{alg:AM} and \ref{alg:PALM} do not satisfy this requirement; they both might end up in saddle points like the origin.
    }
\end{remark}
\begin{algorithm} 
	\caption{\textbf{:}  \textbf{Proximal Alternating Linearized Minimization}} \label{alg:PALM}
	\begin{algorithmic}[1]
		\Require{$\y \in \R^m$, $\AA \in \R^{m\times n_1n_2}$ such that $\A(\U\V^\top) = \AA \cdot \vect(\U\V^\top)$, rank $R$, $\V_0 \in \R^{n_2 \times R}$, $\alpha_1,\alpha_2,\beta_1,\beta_2 > 0$, and sequences of step-sizes $(\lambda_k)_k, (\mu_k)_k \subset (r_-,r_+)$, for $r_+ > r_- > 0$}
		\Statex
		\While{stop condition is not satisfied}
		\Let{$\U_{k+1}$}{$\prox{\tfrac{1}{\lambda_k} \EN_\Alpha} \round{ \U_k - \lambda_k \vect^{-1} \round{ \AA_{\V_k}^\top \round{ \AA_{\V_k} \vect(\U_k) - \y } } }$}
		\Comment{$\A(\U\V^\top) = \AA_\V \cdot \vect(\U)$}
		\Let{$\V_{k+1}$}{$\prox{\tfrac{1}{\mu_k} \EN_\Beta} \round{ \V_k - \mu_k \vect^{-1} \round{ \AA_{\U_{k+1}}^\top \round{ \AA_{\U_{k+1}} \vect(\V_k) - \y } } }$}
		\Comment{$\A(\U\V^\top) = \AA_\U \cdot \vect(\V)$}
		\EndWhile
		\State 
		\Return{$(\U_\text{final},\V_\text{final})$}
	\end{algorithmic}
\end{algorithm}

\section{Proofs} \label{Proofs}

This section provides the remaining proofs for the main results from Section \ref{MainResults}.


\subsection{Proof of Theorem \ref{thm:SmallBallRIP}}
\label{sec:SmallBallRIPProof}
The proof of Theorem \ref{thm:SmallBallRIP} is a straight-forward application of Mendelson's small ball method \cite{mendelson2014learning}. 
We first recap the key result transferred to our setting.
\begin{theorem}[{\cite[Corollary 5.5]{mendelson2014learning}}] \label{thm:Mendelson}
	Let $K \subset \R^{n_1 \times n_2}$ be star-shaped around $\0$, i.e., for any $t \in [0,1]$ and $\Z \in K$ one has $t\Z \in K$. Let $\AA \in \R^{n_1\times n_2}$ be an isotropic random matrix satisfying
	\begin{align*}
		Q_K(2\tau) := \inf_{\Z \in K} \; \Pr{ \abs{\inner{\AA,\Z}{F}}{} \ge 2\tau \norm{\Z}{F} } 
		> 0,
	\end{align*}
	and let $\AA_i$, $i \in [m]$, be i.i.d.\ copies of $\AA$. Let $r > 0$ be sufficiently large to have
	\begin{align*}
		\E{ \sup_{\Z \in K \cap r B_F} \abs{ \inner{ \frac{1}{m} \sum_{i = 1}^m \eps_i \AA_i, \Z }{F} }{} } \le \frac{\tau}{16} Q_K(2\tau) r,
	\end{align*}
	where $B_F$ denotes the Frobenius unit ball. Then with probability at least $1 - 2e^{-\frac{1}{2} Q_K(2\tau)^2 m}$
	\begin{align*}
		\abs{ \curly{ i \colon \abs{ \inner{\AA_i,\Z}{F} }{} \ge \tau \norm{\Z}{F} } }{} \ge \frac{1}{4} Q_K(2\tau) m,
	\end{align*}
    for all $\Z \in K$ with $\norm{\Z}{F} \ge r$.
\end{theorem}
%
%
%
\begin{Proof}[of Theorem \ref{thm:SmallBallRIP}]
	Let us define $K = \Gamma K_{s_1,s_2}^R - \Gamma K_{s_1,s_2}^R$. First note that $K$ is star-shaped and since \eqref{eq:SmallBall} holds for $\AA$, we have $Q_K(2\tau) \ge c$, where $2\tau = \theta$. By \eqref{eq:measSmallBall} and symmetry of $K$, we get that
	\begin{align*}
		\E{ \sup_{\Z \in K \cap r B_F} \abs{ \inner{ \frac{1}{m} \sum_{i = 1}^m \eps_i \AA_i, \Z }{F} }{} } 
		&= \E{ \sup_{\Z \in K \cap r B_F} \inner{ \frac{1}{m} \sum_{i = 1}^m \eps_i \AA_i, \Z }{F} } \\
		&= \frac{E_m(K,\AA)}{\sqrt{m}} \le \frac{\tau}{16} Q_K(2\tau) \delta.
	\end{align*}
	We may now apply Theorem \ref{thm:Mendelson} with $\tau = \frac{\theta}{2}$ and $r = \delta$ to obtain with probability at least $1 - 2e^{\frac{c^2}{2} m}$
	\begin{align*}
		\norm{\A(\Z)}{2}^2
		&= \sum_{i=1}^m \inner{ \frac{1}{\sqrt{m}} \AA_i, \Z }{F}^2 
		= \frac{1}{m} \sum_{i=1}^m \inner{ \AA_i, \Z }{F}^2 \\
		&\ge \frac{1}{m} \cdot \round{ \frac{1}{4} Q_K(2\tau) m} \cdot \round{\tau^2 \norm{\Z}{F}^2} \\
		&\ge \frac{\theta^2 c}{16} \norm{\Z}{F}^2,
	\end{align*}
	for all $\Z \in K$ with $\norm{\Z}{F} \ge \delta$. Since $\Gamma K_{s_1,s_2}^R \subset K$, the claim follows.
\end{Proof}


\subsection{Proof of Corollary \ref{GaussianRIP}}\label{sec:coveringnumbers}

To prove Corollary \ref{GaussianRIP}, we need to bound the complexity of $\Gamma K_{s_1,s_2}^R$ in terms of the Gaussian width $w(\Gamma K_{s_1,s_2}^R)$. Since the Gaussian width of a set $K$ is strongly connected to the covering numbers $N(K,\norm{\cdot}{2},\eps)$ of $K$ with respect to the Euclidean norm, we begin by bounding this quantity.

\begin{lemma}[Covering numbers of $\Gamma K_{s_1,s_2}^R$] \label{CoveringNumber2}
	Let $\Gamma > 0$ and let $K_{s_1,s_2}^{R}$ be the set defined in \eqref{eq:K}. Then, for all $\eps \in (0,1)$, one has that
	\begin{align} \label{CoveringCardinality2}
	\log \round{ N(\Gamma K_{s_1,s_2}^{R}, \Vert \cdot \Vert_F, \eps) } \lesssim \eps^{-2} \Gamma^2 R^3 (s_1 + s_2) \log \round{ \max \curly{ \frac{en_1}{s_1}, \frac{en_2}{s_2} } }.
	\end{align}
\end{lemma}
\begin{Proof}
	First, note that by \cite[Lemma 3.4]{Plan2013LP} the covering number of the set of effectively $s$-sparse vectors $K_{n,s}$ can be bounded by
	\begin{align*}
		\log \round{ N( K_{n,s}, \norm{\cdot}{2}, \eps ) } \lesssim \eps^{-2} s \log \round{ \frac{en}{s} }.
	\end{align*}
	We now construct a net for $K_{s_1,s_2}^R$. Let $\tilde{K}_{Rn_1,Rs_1}$ and $\tilde{K}_{Rn_2,Rs_2}$ be minimal $\frac{\eps}{2 \sqrt{R}}$-nets of $\sqrt{R} K_{Rn_1,Rs_1}$ and $\sqrt{R} K_{Rn_2,Rs_2}$, and define
	\begin{align*}
		\tilde{K} = \curly{ \tilde{\Z} = \tilde{\U} \tilde{\V}^\top \colon \vect(\tilde{\U}) \in \tilde{K}_{Rn_1,Rs_1} \text{ and } \vect(\tilde{\V}) \in \tilde{K}_{Rn_2,Rs_2} }.
	\end{align*}
	Hence, for any $\Z = \U\V^\top \in K_{s_1,s_2}^R$ there exists $\tilde{\Z} = \tilde{\U} \tilde{\V}^\top \in \tilde{K}$ such that $\| \U - \tilde{\U}\|_{F} \le \frac{\eps}{2\sqrt{R}}$ and $\| \V - \tilde{\V} \|_{F} \le \frac{\eps}{2\sqrt{R}}$. This implies
	\begin{align*}
		\| \Z - \tilde{\Z} \|_{F} \le \| \U - \tilde{\U} \|_{F} \norm{\V}{F} + \|\tilde{\U} \|_{F} \| \V - \tilde{\V} \|_{F} \le \frac{\eps}{2\sqrt{R}} \sqrt{R} + \sqrt{R} \frac{\eps}{2\sqrt{R}} = \eps,
	\end{align*}
	i.e., $\tilde{K}$ is an $\eps$-net of $K_{s_1,s_2}^R$. By construction, the cardinality of $\tilde{K}$ is bounded by $| \tilde{K} |~\le~| \tilde{K}_{Rn_1,Rs_1} |~\cdot~| \tilde{K}_{Rn_2,Rs_2}|$ and, consequently, using that $N(cK,\norm{\cdot}{}, \eps) = N \round{ K, \norm{\cdot}{}, \frac{\eps}{c} }$, for $c > 0$, we get
	\begin{align*}
		\log \round{ N(K_{s_1,s_2}^{R}, \Vert \cdot \Vert_F, \eps) } 
		&\le \log \round{ N \round{ \sqrt{R} K_{Rn_1,Rs_1}, \norm{\cdot}{2}, \frac{\eps}{2\sqrt{R}} } } + \log \round{ N \round{ \sqrt{R} K_{Rn_2,Rs_2}, \norm{\cdot}{2}, \frac{\eps}{2\sqrt{R}} } } \\
		&\lesssim \eps^{-2} R^3 (s_1 + s_2) \log \round{ \max \curly{ \frac{en_1}{s_1}, \frac{en_2}{s_2} } }.
	\end{align*}
	The claim follows from $N(\Gamma K_{s_1,s_2}^{R}, \Vert \cdot \Vert_F, \eps) = N \round{ K_{s_1,s_2}^{R}, \Vert \cdot \Vert_F, \frac{\eps}{\Gamma} }$.
\end{Proof}

From Lemma \ref{CoveringNumber2}, we deduce the following bound on the Gaussian width of $K_{s_1,s_2}^R$.

\begin{lemma}[Gaussian width of $\Gamma K_{s_1,s_2}^R$] \label{lem:GaussianWidth}
	Let $\Gamma > 0$ and let $K_{s_1,s_2}^{R}$ be the set defined in Section \ref{MainResults}. Then,
	\begin{align*}
		w(\Gamma K_{s_1,s_2}^R) \lesssim \sqrt{ \Gamma^2 R^3 (s_1 + s_2) \log^3 \round{ \max \curly{n_1, n_2} } }.
	\end{align*}
\end{lemma}

\begin{Proof}
	The two-sided Sudakov inequality \cite[Theorem 8.1.13]{vershynin2018high} yields that
	\begin{align*}
		w(\Gamma K_{s_1,s_2}^R) \lesssim \log(n_1 n_2) \; \sup_{\eps > 0} \; \eps \sqrt{ \log \round{ N( \Gamma K_{s_1,s_2}^R, \norm{\cdot}{F}, \eps ) } }.
	\end{align*}
	Lemma \ref{CoveringNumber2} now yields the claim.
\end{Proof}

\rev{
\begin{Proof}[of Corollary \ref{GaussianRIP}]
    We assume that $\AA \in R^{n_1 \times n_2}$ is isotropic and $\mathcal K$-subgaussian. First note that, for any set $K \subset \R^{n_1 \times n_2}$, the mean empirical width $E_m(K,\AA)$ can be bounded by the Gaussian width $w(K)$ \cite[Eq.\ (6.4)]{tropp2015convex} in this case. Hence,
	\begin{align*}
		E_m (\Gamma K_{s_1,s_2}^R - \Gamma K_{s_1,s_2}^R, \AA)^2 
		&\le C w(\Gamma K_{s_1,s_2}^R - \Gamma K_{s_1,s_2}^R)^2 
		\le 4C w(\Gamma K_{s_1,s_2}^R)^2 \\
		&\lesssim \Gamma^2 R^3 (s_1 + s_2) \log^3 \round{ \max \curly{n_1, n_2} },
	\end{align*} 
	where the second step follows from elementary properties of $w$ and the third step from Lemma \ref{lem:GaussianWidth}. Consequently, \eqref{eq:measSmallBall} becomes
	\begin{align} \label{eq:HeavyTailed}
		m \gtrsim \delta^{-2} \Gamma^2 R^3 (s_1 + s_2) \log^3 \round{ \max \curly{n_1, n_2} }.
	\end{align}
	All that remains is to show that $\AA$ satisfies \eqref{eq:SmallBall} on $\Gamma K_{s_1,s_2}^R - \Gamma K_{s_1,s_2}^R$ with $c$ only depending on $\theta$ and $\mathcal K$. To this end, recall from Section \ref{sec:SubgaussianDetails} that for any $\Z \in \R^{n_1\times n_2}$ we have by assumption that $\| \inner{\AA,\Z}{F} \|_{L_2} = \| \Z \|_F$ and $\| \inner{\AA,\Z}{F} \|_{L_p} \lesssim \mathcal K \sqrt{p} \| \inner{\AA,\Z}{F} \|_{L_2}$, for any $p \ge 1$ ($\AA$ is isotropic and subgaussian). Note that by Hölder's and Jensen's inequalities the second property extends to 
    \begin{align} \label{eq:GaussianProof}
        \| \inner{\AA,\Z}{F}^2 \|_{L_p} \le \| \inner{\AA,\Z}{F} \|_{L_{2p}}^2 
        \lesssim \mathcal K^2 p \| \inner{\AA,\Z}{F} \|_{L_2}^2
        \le \mathcal K^2 p \| \inner{\AA,\Z}{F}^2 \|_{L_2}
    \end{align}
    By the Paley-Zygmund inequality \cite[Corollary 3.3.2]{de2012decoupling}, we thus obtain
	\begin{align*}
	    \Pr{\inner{\AA,\Z}{F}^2 \ge \theta \norm{\Z}{F}^2 } &= \Pr{\inner{\AA,\Z}{F}^2 \ge \theta \| \inner{\AA,\Z}{F} \|_{L_2}^2 }
	    \ge \Pr{\inner{\AA,\Z}{F}^2 \ge \theta \| \inner{\AA,\Z}{F}^2 \|_{L_2} } \\
	    &\ge \Big( (1-\theta^2) \frac{ \| \inner{\AA,\Z}{F}^2 \|_{L_2}^2 }{ \| \inner{\AA,\Z}{F}^2 \|_{L_3}^2 } \Big)^3
	    \gtrsim \Big( (1-\theta^2) \frac{1}{\mathcal K^4} \Big)^3
	\end{align*}
	and hence the claim. In the last step, we used \eqref{eq:GaussianProof} for $p = 3$.
\end{Proof}
}

\subsection{Proof of Lemma \ref{lem:Talagrand}}
\label{sec:Rank1measProof}
%

    Define $X_\Z := \inner{\sum_{i=1}^m \eps_i \a_i \a_i^\top, \Z}{F}$, such that $E_m(K,\AA) = \frac{1}{\sqrt{m}} \E{\sup_{\Z \in K} X_\Z}$ and recall that $\eps_1,\dots,\eps_m \in \{-1,1\}$ are i.i.d.\ Bernoulli variables. We show that
    \begin{align} \label{eq:Concentration}
       \Pr{ \abs{X_\Z - X_{\Z'}}{} > t} \le 2e^{-c \min\curly{ \frac{t^2}{m R \norm{\Z-\Z'}{F}^2}, \frac{t}{\norm{\Z-\Z'}{}} }},
    \end{align}
    for $\Z,\Z' \in \R^{n\times n}$ and apply generic chaining to obtain the claim. Note that \eqref{eq:Concentration} is equivalent to showing \rev{for any $\Z$ that}
    \begin{align} \label{eq:Rank1concentration}
        \norm{X_\Z}{L^q} := \E{\abs{X_\Z}{}^q}^{\frac{1}{q}} \lesssim \sqrt{m R} \norm{\Z}{F} \sqrt{q} + \norm{\Z}{} q,
        \quad \text{for all } q \ge 1.
    \end{align}
    \rev{This equivalence follows from $X_\Z - X_{\Z'} = X_{\Z - \Z'}$ and the equivalence of tail bounds and moment growth for subgaussian \cite[Proposition 2.5.2]{vershynin2018high} and subexponential \cite[Proposition 2.7.1]{vershynin2018high} random variables.}
    By using the triangle inequality, we can estimate
    \begin{align} \label{eq:triangle}
    \begin{split}
        \E{\abs{X_\Z}{}^q}^{\frac{1}{q}} 
        &= \E[\eps]{\E[\AA]{\abs{X_\Z}{}^q}}^{\frac{1}{q}}
        = \E[\eps]{ \norm{X_\Z}{L^q,\AA}^q }^{\frac{1}{q}} \\
        &\le \E[\eps]{ \round{ \norm{X_\Z - \E[\AA]{X_\Z}}{L^q,\AA} + \norm{\E[\AA]{X_\Z}}{L^q,\AA} }^q }^{\frac{1}{q}} \\
        &= \norm{ \norm{X_\Z - \E[\AA]{X_\Z}}{L^q,\AA} + \norm{\E[\AA]{X_\Z}}{L^q,\AA} }{L^q,\eps} \\
        &\le \norm{ \norm{X_\Z - \E[\AA]{X_\Z}}{L^q,\AA} }{L^q,\eps} + \norm{ \norm{\E[\AA]{X_\Z}}{L^q,\AA} }{L^q,\eps},
    \end{split}
    \end{align}
    where $\E[\eps]{\cdot}$ and $\E[\AA]{\cdot}$ denote conditional expectations that keep everything fixed apart from the random variable in the subscript, and $\norm{\cdot}{L^q,\eps}$ and $\norm{\cdot}{L^q,\AA}$ denote the $L^q$-norm with respect to $\eps$ resp.\ $\AA$. To estimate the first summand on the right-hand side of \eqref{eq:triangle}, rewrite $X_\Z = \tilde{\a}^\top \MM_{\Z,\eps} \tilde{\a}$ where $\tilde{\a} = (\a_1^\top,...,\a_m^\top)^\top \in \R^{mn}$ is the concatenation of all $\a_i$ and $\MM_{\Z,\eps} \in \R^{mn\times mn}$ is a block diagonal matrix with blocks $\eps_i \Z$. Note that $\norm{\MM_{\Z,\eps}}{F} = \sqrt{m} \norm{\Z}{F}$ and $\norm{\MM_{\Z,\eps}}{} = \norm{\Z}{}$. For $\eps$ and $\Z$ fixed, the Hanson-Wright inequality \cite[Theorem 6.2.1]{vershynin2018high} yields
    \begin{align*}
        \Pr{\abs{ X_\Z - \E[\AA]{X_\Z} }{} > t} \le 2e^{-c \min\curly{ \frac{t^2}{m\norm{\Z}{F}^2}, \frac{t}{\norm{\Z}{}} }},
    \end{align*}
    for $c > 0$ only depending on $\mathcal{K}$, which is equivalent to
    \begin{align} \label{eq:Rank1firstTerm}
        \norm{X_\Z - \E[\AA]{X_\Z}}{L^q,\AA} \lesssim \sqrt{m} \norm{\Z}{F} \sqrt{q} + \norm{\Z}{} q
    \end{align}
    (and independent of $\eps$). To bound the second term, note that
    \rev{
    \begin{align*}
        \E[\AA]{X_\Z} 
        &= \E[\AA]{\tilde{\a}^\top \MM_{\Z,\eps} \tilde{\a}}
        = \E[\AA]{\trace(\tilde{\a}^\top \MM_{\Z,\eps} \tilde{\a})} 
        = \E[\AA]{\trace(\MM_{\Z,\eps} \tilde{\a} \tilde{\a}^\top)} \\
        &= \trace(\MM_{\Z,\eps} \E[\AA]{\tilde{\a} \tilde{\a}^\top})
        = \trace(\MM_{\Z,\eps} \cdot \id) \\
        &= \sum_{i=1}^m \eps_i \trace(\Z),
    \end{align*}
    }
    which implies by Hoeffding's inequality \cite[Theorem 2.2.5]{vershynin2018high}
    \begin{align*}
        \Pr[\eps]{\abs{\E[\AA]{X_\Z}}{} > t} \le 2e^{-\frac{t^2}{2m \trace(\Z)^2}}.
    \end{align*}
    Consequently,
    \begin{align} \label{eq:Rank1secondTerm}
        \norm{ \norm{\E[\AA]{X_\Z}}{L^q,\AA} }{L^q,\eps} 
        = \norm{ \E[\AA]{X_\Z} }{L^q,\eps} \lesssim \sqrt{m} \; \trace(\Z) \sqrt{q}.
    \end{align}
    Combining \eqref{eq:Rank1firstTerm} and \eqref{eq:Rank1secondTerm} and using that $\trace(\Z) \le \norm{\Z}{*} \le \sqrt{R} \norm{\Z}{F}$ yields \eqref{eq:Rank1concentration}. We conclude by applying generic chaining \cite[Theorem 2.2.23]{talagrand2014upper} to obtain
    \begin{align*}
        E_m(K,\AA) = \frac{1}{\sqrt{m}} \E{\sup_{\Z \in K} X_\Z} \le C \round{ \sqrt{R} \gamma_2(K,\norm{\cdot}{F}) + \frac{\gamma_1(K,\norm{\cdot}{})}{\sqrt{m}} }.
    \end{align*}


\rev{

\subsection{Proofs of Theorems \ref{LocalConvergence_AM} and \ref{LocalConvergence_PALM}}
\label{sec:ConvergenceProof}

Let us first verify that $J_\abg^R$ satisfies the basic assumptions $(\mathcal{H})$ and $(\mathcal{H}1)$ in \cite{attouch2010proximal} resp.\ Assumptions 1 and 2 in \cite{bolte2014proximal}. For convenience, we state them in our notation and re-group them in the following way:
\begin{align*}
(\mathcal H) \tab 
&\begin{cases}
J_\abg^R (\U,\V) = \underset{ f(\U)}{\underbrace{\Big( \alpha_1 \norm{\U}{F}^2 + \alpha_2 \norm{\U}{1} \Big)}} + \underset{ Q(\U,\V)}{\underbrace{\left\Vert \y - \mathcal A \left( \U\V^\top \right) \right\Vert_2^2}} + \underset{g(\V)}{\underbrace{\Big( \beta_1 \norm{\V}{F}^2 + \beta_2 \norm{\V}{1} \Big)}}, \\
f: \mathbb{R}^{n_1 \times R} \rightarrow {\mathbb{R} \cup \{ \infty \}}, \; g:\mathbb{R}^{n_2 \times R} \rightarrow {\mathbb{R} \cup \{ \infty \}} 
\text{ are proper lower semicontinuous}, \\
Q: \mathbb{R}^{n_1 \times R} \times \mathbb{R}^{n_2 \times R}  \rightarrow \mathbb{R} \text{ is a }
C^1 \text{ function}, \\
\inf J_\abg^R > -\infty , \; \inf f > -\infty , \text{ and } \inf g > -\infty , \\
\nabla Q \text{ is Lipschitz continuous on bounded subsets of } \mathbb{R}^{n_1 \times R} \times \mathbb{R}^{n_2 \times R}, \\
\nabla_\U Q \text{ is globally } L_1(\V) \text{-Lipschitz, for fixed } \V, \\
\nabla_\V Q \text{ is globally } L_2(\U) \text{-Lipschitz, for fixed } \U.
\end{cases}
\end{align*}
It is straight-forward to check that $(\mathcal H)$ is satisfied for $J_\abg^R$. In addition, the sequence of iterates $(\U_k,\V_k)$ generated by Algorithms \ref{alg:AM} and \ref{alg:PALM} shall satisfy
\begin{align*}
	(\mathcal H_1) \tab \text{there exists } \lambda_+,\lambda_- > 0 \text{ such that } \lambda_- \le L_1(\U_k), \; L_2(\V_k) \le \lambda_+ \text{, for all } k \in \mathbb{N}.
\end{align*}
This is trivially fulfilled if $(\U_k,\V_k)$ stays bounded. Since $J_\abg^R$ is coercive and both algorithms are descent methods, the sequence $(\U_k,\V_k)$ will not diverge to $\infty$ and $(\mathcal H_1)$ holds.

Let us now validate that $J_\abg^R$ has the KL-property. To do so, we show that $J_\abg^R$ is a semialgebraic function since semialgebraic functions satisfy the KL-property at each point with $\varphi(t) = ct^{1-\theta}$ for some $\theta \in [0,1) \cap \mathbb{Q}$ and $c > 0$, see \cite[Section 4.3]{attouch2010proximal}. Using this strategy, we however pay the price of having no better knowledge on the parameters $\eps$ and $\eta$ in Theorems \ref{LocalConvergence_AM} and \ref{LocalConvergence_PALM}, which characterize the convergence radius. A function $f \colon \R^d \to \R$ is semialgebraic if $\graph(f) \subset \mathbb{R}^{d} \times \mathbb{R}$ is a semialgebraic set; a set in $\mathbb{R}^d$ is called semialgebraic if it can be written as a finite union of sets of the form
\begin{align*}
\{ x \in \mathbb{R}^d \; : \; p_i(x) = 0, \; q_i(x) \purple{>} 0, \; i = 1,\dots,p \},
\end{align*}
where $p_i,q_i$ are real polynomials. Clearly, the absolute value of a single entry of a vector $h(x) := \vert x_l \vert$ is a semialgebraic function as
\begin{align*}
\graph (h) = \{ (x,r) \in \mathbb{R}^d \times \mathbb{R} : x_i + r = 0, \; -x_i > 0 \}
\cup \{ (x,r) \in \mathbb{R}^d \times \mathbb{R} : x_i = 0, \; r = 0 \} \\
\cup \{ (x,r) \in \mathbb{R}^d \times \mathbb{R} : x_i - r = 0, \; x_i > 0 \}.
\end{align*}
Similarly, polynomials $p$ are semialgebraic as $\graph (p) = \{ (x,r) \in \mathbb{R}^d \times \mathbb{R} : p(x) - r = 0 \}$. Finally, composition, finite sums and finite products of semialgebraic functions are semialgebraic. The semialgebraicity of $J_\abg^R$ follows as
\begin{align*}
J_\abg^R(U,V)
= \sum_{l = 1}^m \vert y_l - \langle A_l , UV^\top 
\rangle_F \vert^2 &+ \alpha_1  \sum_{i = 1}^{n_1} \sum_{r = 1}^R  \vert U_{i,r} \vert^2 + \alpha_2  \sum_{i = 1}^{n_1} \sum_{r = 1}^R  \vert U_{i,r} \vert \\
&+ \beta_1  \sum_{i = 1}^{n_2} \sum_{r = 1}^R  \vert V_{i,r} \vert^2 + \beta_2  \sum_{i = 1}^{n_2} \sum_{r = 1}^R  \vert V_{i,r} \vert
\end{align*}
is just a finite composition of semialgebraic basic units. Consequently, the functional $J_\abg^R$ has the KL-property and hence meets all basic requirements in \cite{attouch2010proximal,bolte2014proximal}.

Whereas this suffices to deduce Theorem \ref{LocalConvergence_PALM} from the results in \cite{bolte2014proximal}, it requires some additional work to see that the claim of \rev{\cite[Theorem 3.3]{attouch2010proximal}} is still valid if the sequence $(\U_k,\V_k)$ is defined via alternating minimization in Algorithm \ref{alg:AM} and not proximal alternating minimization as in \cite{attouch2010proximal}. We first make the following observation, which replaces \cite[Lemma 3.1]{attouch2010proximal}.

\begin{lemma} \label{lem:Lemma5}
    For $J_\abg^R$ and $(\U_k,\V_k)$ defined by Algorithm \ref{alg:AM}, we have that
    \begin{align*}
        J_\abg^R(\U_k,\V_k) - J_\abg^R(\U_{k+1},\V_{k+1}) \ge \alpha_1 \norm{\U_k - \U_{k+1}}{F}^2 + \beta_1 \norm{\V_k - \V_{k+1}}{F}^2,
    \end{align*}{}
    and
    \begin{align*}
        \sum_{k=0}^\infty \round{ \norm{\U_k - \U_{k+1}}{F}^2 + \norm{\V_k - \V_{k+1}}{F}^2 } < \infty,
    \end{align*}{}
    which implies that $\lim_{k \rightarrow \infty} \round{ \norm{\U_k - \U_{k+1}}{F}^2 + \norm{\V_k - \V_{k+1}}{F}^2 } = 0$. Moreover,
    \begin{align*}
        2 \begin{pmatrix}
            \A_{\V_k}^* \round{ \A(\U_k \V_k^\top) - \y } - \A_{\V_{k-1}}^* \round{ \A(\U_k \V_{k-1}^\top) - \y } \\
            0
        \end{pmatrix}
        \in \partial J_\abg^R(\U_k,\V_k),
    \end{align*}{}
    where $\A_\V \colon \R^{n_1\times R} \rightarrow \R^m$ is defined such that $\A_\V(\U) = \A(\U\V^\top)$.
\end{lemma}{}

\begin{Proof}
    First note that $J_\abg^R(\cdot,\V)$ is $2\alpha_1$-strongly convex and $J_\abg^R(\U,\cdot)$ is $2\beta_1$-strongly convex, for all $\U \in \R^{n_1 \times R}, \V \in \R^{n_2 \times R}$. To see this, check that $J_\abg^R(\cdot,\V) - \alpha_1 \norm{\cdot}{F}^2$ and $J_\abg^R(\U,\cdot) - \beta_1 \norm{\cdot}{F}^2$ are convex. Since $\U_{k+1}$ minimizes $J_\abg^R(\cdot,\V_k)$, we know that $\0 \in \partial J_\abg^R(\U_{k+1},\V_k)$. Since in addition $\partial [\alpha_1 \| \cdot \|_F^2](\U) = 2\alpha_1 \U$, for any $\U \in \R^{n_1 \times R}$, we deduce that
    \begin{align*}
        -2\alpha_1 \U_{k+1} \in \partial [J_\abg^R(\cdot,\V_k) - \alpha_1 \norm{\cdot}{F}^2] (\U_{k+1}).
    \end{align*}
    By convexity of $J_\abg^R(\cdot,\V) - \alpha_1 \norm{\cdot}{F}^2$, we hence get that
    \begin{align*}
        J_\abg^R(\U_k,\V_k) - \alpha_1 \norm{\U_k}{F}^2 
        \ge (J_\abg^R(\U_{k+1},\V_k) - \alpha_1 \norm{\U_{k+1}}{F}^2) - 2\alpha_1 \inner{\U_{k+1},\U_k - \U_{k+1}}{},
    \end{align*}
    which implies by rearranging
    \begin{align*}
        J_\abg^R(\U_k,\V_k) - J_\abg^R(\U_{k+1},\V_k)
        \ge \alpha_1 \norm{\U_k - \U_{k+1}}{F}^2.
    \end{align*}
    The same argument applied to $\V_{k+1}$ yields
    \begin{align*}
        J_\abg^R(\U_{k+1},\V_k) - J_\abg^R(\U_{k+1},\V_{k+1})
        \ge \beta_1 \norm{\V_k - \V_{k+1}}{F}^2.
    \end{align*}
    We obtain
    \begin{align*}
        J_\abg^R(\U_k,\V_k) 
        &\ge J_\abg^R(\U_{k+1},\V_k) + \alpha_1 \norm{\U_k - \U_{k+1}}{F}^2 \\
        &\ge J_\abg^R(\U_{k+1},\V_{k+1}) + \alpha_1 \norm{\U_k - \U_{k+1}}{F}^2 + \beta_1 \norm{\V_k - \V_{k+1}}{F}^2
    \end{align*}{}
    and hence the first claim. The second claim directly follows by
    \begin{align*}
        \sum_{k=0}^\infty \round{ \norm{\U_k - \U_{k+1}}{F}^2 + \norm{\V_k - \V_{k+1}}{F}^2 } &\le \frac{1}{\min\curly{\alpha_1,\beta_1}} \sum_{k=0}^\infty \round{ \alpha_1 \norm{\U_k - \U_{k+1}}{F}^2 + \beta_1 \norm{\V_k - \V_{k+1}}{F}^2 } \\
        &\le \frac{1}{\min\curly{\alpha_1,\beta_1}} \sum_{k=0}^\infty \round{ J_\abg^R(\U_k,\V_k) - J_\abg^R(\U_{k+1},\V_{k+1}) } \\
        &= \frac{1}{\min\curly{\alpha_1,\beta_1}} J_\abg^R(\U_0,\V_0) < \infty.
    \end{align*}
    Let us now turn to the last claim. By minimality of $\U_{k+1}$ and $\V_{k+1}$, we know that
    \begin{align*}
        \0 \in \partial_\U J_\abg^R(\U_{k+1},\V_k) \quad \text{and} \quad \0 \in \partial_\V J_\abg^R(\U_{k+1},\V_{k+1}).
    \end{align*}{}
    Since $\partial_\U J_\abg^R(\U,\V) = 2\A_\V^* \round{ \A(\U\V^\top) - \y } + 2\alpha_1 \U + \alpha_2 \sign(\U)$, where $\sign$ is applied component-wise, we have that
    \begin{align*}
        - 2\A_{\V_k}^* \round{ \A(\U_{k+1} \V_k^\top) - \y } - 2\alpha_1 \U_{k+1} \in \alpha_2 \sign(\U_{k+1})
    \end{align*}
    and thus
    \begin{align*}
        2 \A_{\V_{k+1}}^* \round{ \A(\U_{k+1} \V_{k+1}^\top) - \y } - 2 \A_{\V_{k}}^* \round{ \A(\U_{k+1} \V_{k}^\top) - \y } &\in \partial_\U J_\abg^R(\U_{k+1},\V_{k+1}), \\
        \0 &\in \partial_\V J_\abg^R(\U_{k+1},\V_{k+1}).
    \end{align*}{}
\end{Proof}{}

With the help of Lemma \ref{lem:Lemma5} it is straight-forward to show that \cite[Proposition 3.1]{attouch2010proximal} holds for $J_\abg^R$ with $(\U_k,\V_k)$ defined by Algorithm \ref{alg:AM}. Combining Lemma \ref{lem:Lemma5} and the adapted version of \cite[Proposition 3.1]{attouch2010proximal}, the pre-convergence result \cite[Theorem 3.1]{attouch2010proximal} holds as well for $J_\abg^R$ with $(\U_k,\V_k)$ defined by Algorithm \ref{alg:AM}. Theorem \ref{LocalConvergence_AM} is a direct consequence, cf.\ \cite[Theorems 3.2 \& 3.3]{attouch2010proximal}.
}


\section{Numerical Experiments} \label{Numerics}

We finally compare our theoretical predictions to the empirical performance 
of our alternating methods. Although Algorithm \ref{alg:AM} is the less efficient implementation (in each alternating step one has to compute a full proximal gradient descent), we prefer it in our simulations. It is not as sensitive to step size adaption as Algorithm \ref{alg:PALM} and thus diminishes the need of parameter tuning, cf.\ Section \ref{sec:NumericsAlg1vsAlg3}.\\
First, we check if the main theoretical result stated in Theorem \ref{ApproxX} describes the qualitative and quantitative behavior of the approximation error well. Second, we demonstrate the performance of Algorithm \ref{alg:AM} on different measurement ensembles, comparing reconstruction for Gaussian, log-normal, and Gaussian rank-1 type measurements. \rev{Then, we examine the runtime of Algorithms \ref{alg:AM} and \ref{alg:PALM} and the influence of initialization quality on the reconstruction accuracy.} Finally, we compare our method to the initially mentioned Sparse Power Factorization (SPF) \cite{lee2013near} which will serve as a general benchmark. In \cite{lee2013near}, SPF has been shown to outperform conventional reconstruction methods that solely rely on low-rankness or sparsity.

\paragraph{Ground-truth.} In order to produce effectively sparse random samples $\X_\star = \U_\star \V_\star^\top \in K_{s_1,s_2}^{R}$, we draw for $\U_\star$ resp.\ $\V_\star$ randomly $Rs_1$ resp.\ $Rs_2$ positions in a $n_1 \times R$- resp.\ $n_2 \times R$-zero-matrix, fill them with Gaussian i.i.d.\ entries, re-normalize to Frobenius norm $\sqrt{R}$ and add a dense Gaussian random matrix of size $n_1 \times R$ resp.\ $n_2 \times R$ of Frobenius norm $0.1 \sqrt{R}$. 

\paragraph{Measuring effective sparsity.} \rev{Given a matrix $\W \in \R^{n\times R}$, for $n = n_1, n_2$, we measure its effective sparsity as $\tfrac{\| \W \|_1^2}{R\| \W \|_F^2} \in [1,\tfrac{\| \W \|_0}{R}]$ and its relative effective sparsity as $\tfrac{\| \W \|_1^2}{(nR) \| \W \|_F^2} \in [0,1]$. These definitions correspond to the average effective sparsity of the columns of $\W$ and are motivated by the observation that any $\Z \in \R^{n_1\times n_2}$ with decomposition $\Z = \U\V^\top$, for $\U \in \R^{n_1 \times R}$ and $\V \in \R^{n_2 \times R}$ with $\tfrac{\| \U \|_1^2}{R \| \U \|_F^2} = s_\U$ and $\tfrac{\| \V \|_1^2}{R \| \V \|_F^2} = s_\V$, satisfies $\Z \in \Gamma K_{\lceil s_\U \rceil, \lceil s_\V \rceil}^R$, for $\Gamma = \tfrac{\| \U \|_F \| \V \|_F}{R}$. (Note that if $\Z$ is of unit norm and $\U\V^\top$ is an orthogonal decomposition, then $\U$ and $\V$ can be rescaled to have both norm $\| \Z \|_F^{\frac{1}{2}}$ such that $\Gamma = \tfrac{1}{R}$ as in the discussion in Remark \ref{rem:dim}.) }

\paragraph{Initialization.} \rev{We use two types of initialization in our experiments: (i) spectral initialization, i.e., $\U_0$ and $\V_0$ are initialized as the leading $R$ left and right singular vectors of $\A^*(\y)$, and (ii) a randomly perturbed ground-truth, i.e., we randomly perturb $\U_\star$ and $\V_\star$ to obtain $\U_0 = \U_\star + \tilde\U$ and $\V_0 = \V_\star + \tilde\V$, where the entries of $\tilde\U, \tilde\V$ are drawn uniformly at random from $[0,0.4 \| \U_\star \|_F]$ and $[0,0.4 \| \V_\star \|_F]$ respectively. This yields an initial error of $\| \X_\star - \X_0 \|_{F} = \| \U_\star \V_\star^\top - \U_0 \V_0^\top \|_{F} \approx 0.6 \| \X_\star \|_{F}$. Whereas (i) represents a practical initialization method that works heuristically in many cases, (ii) is well-suited to validate our theoretical insights. Consequently, we use (ii) in the synthetic examples in Sections \ref{sec:Numerics1}-\ref{sec:Init}, and (i) in the image recovery experiment in Section~\ref{Numerics3}. } 

\paragraph{Parameter tuning.} Since the first experiment in Section \ref{sec:Numerics1} suggests that Algorithm \ref{alg:AM} with $\alpha_1 = \alpha_2 = \beta_1 = \beta_2$ performs slightly better than with the parameter ratio given in Theorem \ref{ApproxX}, in all remaining experiments we set $\alpha_1 = \alpha_2 = \beta_1 = \beta_2 = \mu$ and loosely tune $\mu$ by the heuristic proposed at the end of Section \ref{sec:RecoveryPropertiesRIP}:
\begin{enumerate}
    \item Initialize $\mu$ sufficiently large to guarantee $\X_\abg = \0$, i.e., $\| \X_\star - \X_\abg \|_F / \| \X_\star \|_F = 1$
    \item Iteratively shrink $\mu$ by a factor $\tfrac{1}{2}$ until the relative error $\| \X_\star - \X_\abg \|_{F} / \| \X_\star \|_{F}$ stops decreasing.
\end{enumerate}
Only in Section \ref{Numerics3} we run the simulation on a fixed grid of $\abg$-values to be able to compare parameter tuning via best approximation with parameter tuning via discrepancy principle, cf.\ Table \ref{fig:Table1}.

\subsection{Validation of Theorem \ref{ApproxX}} 
\label{sec:Numerics1}

\begin{figure}[ht!]
 	\centering
 	\captionsetup{width=0.95\linewidth}
 	\begin{subfigure}[b]{0.45\textwidth}
 		\includegraphics[width=\textwidth]{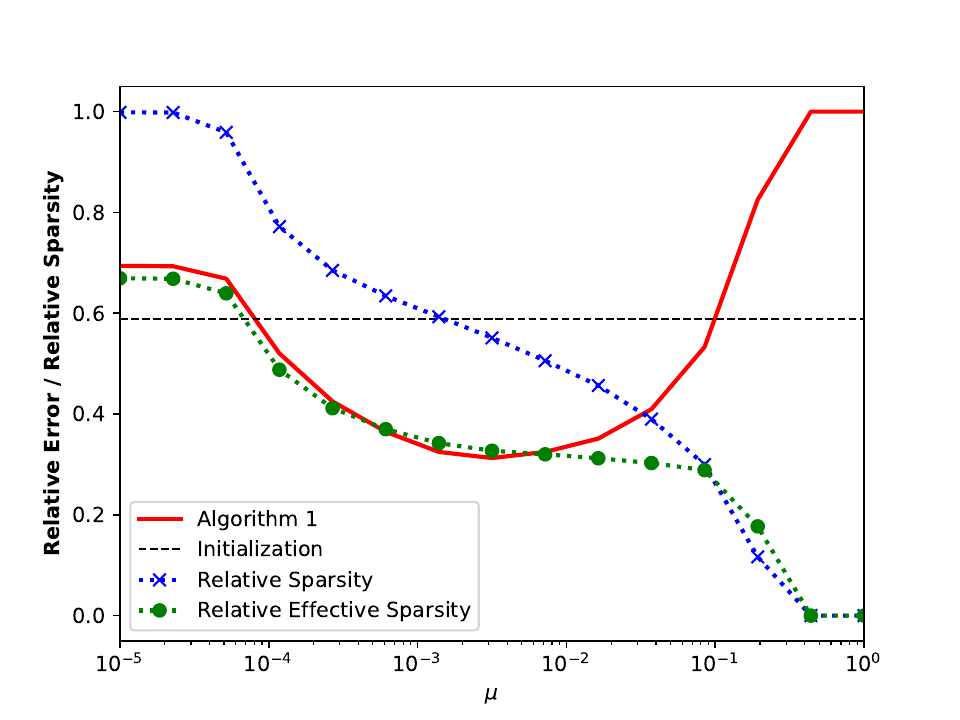}
 		\subcaption{$\alpha_1 = \sqrt{20} \mu = \beta_1 = \sqrt{20} \beta_2$.}
 	\end{subfigure}
 	\quad
 	\begin{subfigure}[b]{0.45\textwidth}
 		\includegraphics[width=\textwidth]{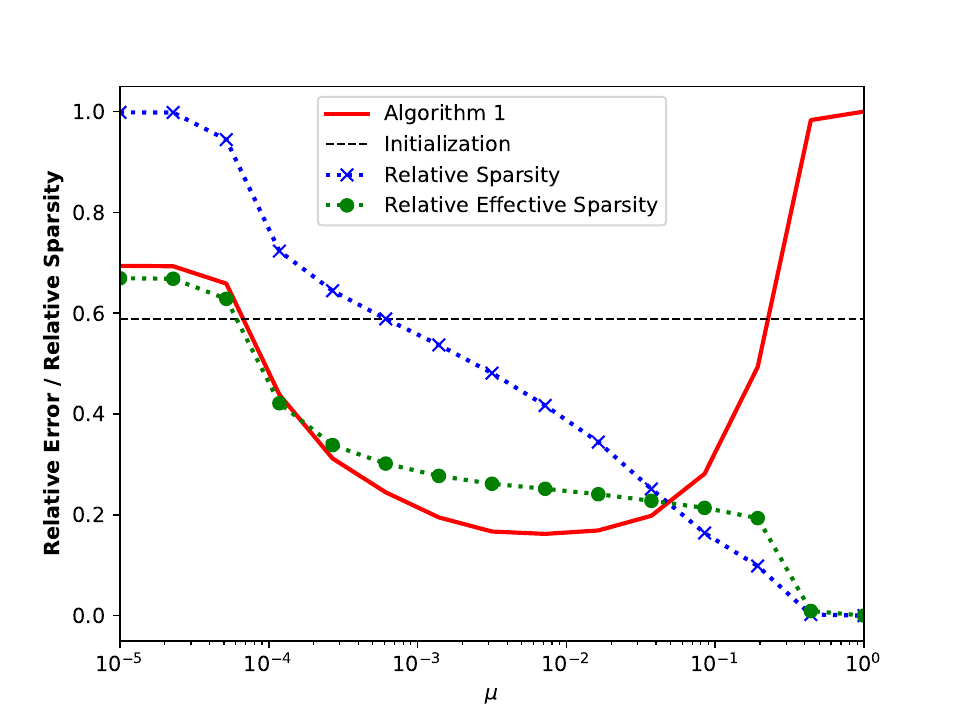}
 		\subcaption{$\alpha_1 = \mu = \beta_1 = \beta_2$.}
 	\end{subfigure}
  \caption{Approximation quality and sparsity depending on parameter size. The approximation error is measured relative to $\Vert \X_\star \Vert_F$ while the sparsity and effective sparsity of $\V_\text{final}$ are measured relative to $\mathrm{rank}(\X_\star) \cdot n_2$, cf.\ Section \ref{Numerics}. For comparison, the approximation error of the initialization is added.}
\label{fig:ParameterTest}
\end{figure}

In the first experiment we study the influence of the parameters $\abg$ on the reconstruction accuracy. Figure~\ref{fig:ParameterTest} shows, first, the average relative error in reconstructing $100$ randomly drawn $\X_\star \in K_{20,20}^1 \subset \R^{20 \times 300}$ from $m=160$ measurements with $\| \Eta \|_{F} = 0.05 \| \X_\star \|_{F}$ and, second, the corresponding (effective) sparsity level measured relative to $n_2 = 300$, i.e., $\frac{s}{n_2} \in [0,1]$, cf.\ beginning of Section \ref{Numerics}. In particular, we set $\alpha_2 =  \mu$ and compare the performance of Algorithm \ref{alg:AM} for $\alpha_1 = \sqrt{20} \mu = \beta_1 = \sqrt{20} \beta_2$ (as proposed in Lemma \ref{Bound-uv2}) with $\alpha_1 = \mu = \beta_1 = \beta_2$. Since we are interested in validating the theory, we only use the randomly perturbed ground-truth here, see Section \ref{Numerics}. \\
The behavior is as predicted by Theorem \ref{ApproxX}: decreasing the parameter(s) shrinks the reconstruction error up to a small multiple of the noise level while the sparsity stays under control, cf.\ Lemma \ref{Bound-uv2}. As soon as the noise level is hit (to be precise, a slightly higher value caused by the injectivity constants $\gamma$ and $\delta$ in \eqref{ApproxX}), the assumptions of Lemma \ref{Bound-uv2} fail, the regularity of the solution vanishes, and the approximation guarantee breaks. Note that setting all parameters $\alpha_1,\alpha_2,\beta_1,\beta_2$ to the same value performs better than the theoretically motivated choice so we keep this in the remaining simulations. Furthermore, let us mention that sparse reconstruction (information theoretic lower bound: $m \ge s_1 s_2 = 400$) and low-rank matrix sensing (information theoretic lower-bound: $m \ge \mathrm{rank}(\X_\star) (n_1 + n_2) = 320$) would not allow reliable reconstruction of $\X_\star$ in this setting.\\
As a byproduct, the experiment suggests the simple parameter choice heuristic of starting with large $\abg$ and shrinking the parameters until there is a drastic change in regularity of the solution. 

\subsection{Validation of Theorem \ref{thm:SmallBallRIP} and Corollary \ref{GaussianRIP} }
\label{sec:Numerics2}

\begin{figure}[ht!]
 	\centering
 	\captionsetup{width=0.95\linewidth}
 	\begin{subfigure}[b]{0.45\textwidth}
 		\includegraphics[width=\textwidth]{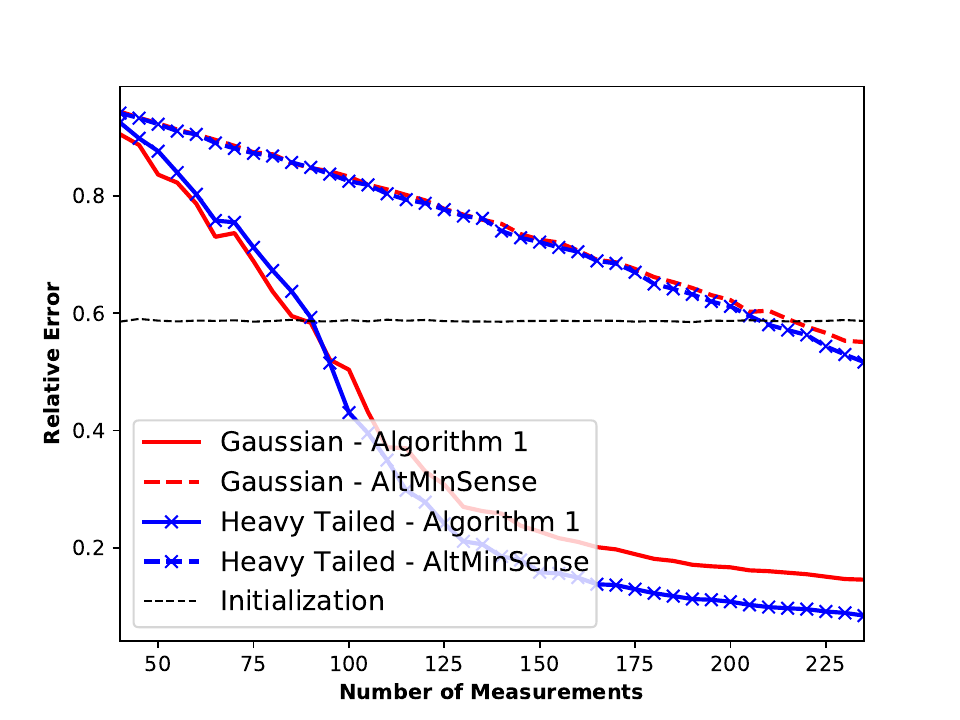}
 		\subcaption{Gaussian and Heavy-tailed measurements.}
 		\label{fig:MeasurementTestA}
 	\end{subfigure}
 	\quad
 	\begin{subfigure}[b]{0.45\textwidth}
 		\includegraphics[width=\textwidth]{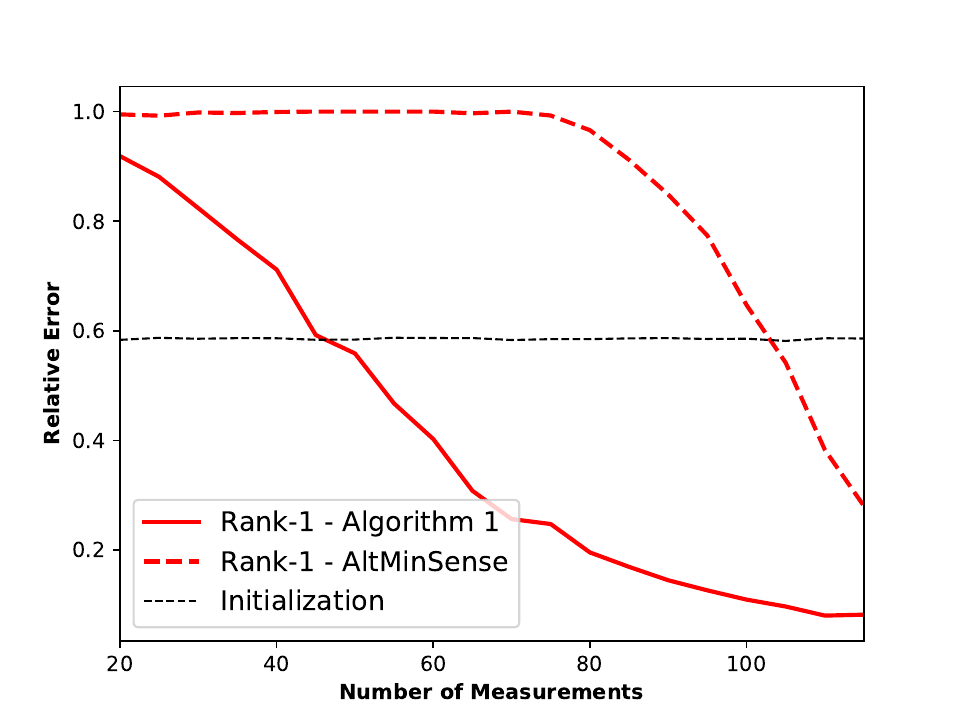}
 		\subcaption{Gaussian rank-$1$ measurements.}
 		\label{fig:MeasurementTestB}
 	\end{subfigure}
  \caption{Approximation quality for various choices of $\A$. The approximation error is measured relative to $\Vert \X_\star \Vert_F$. For comparison, the state-of-the-art low-rank matrix sensing algorithm AltMinSense is added. Note that the information theoretic lower bound for guaranteed reconstruction of matrices only using low-rank structure is $m \ge \rank(\X_\star) (n_1 + n_2) = 320$ in (a) and $m \ge 100$ in (b).}
\label{fig:MeasurementTest}
\end{figure}

In a second experiment we compare the reconstruction performance of Algorithm \ref{alg:AM} with respect to different measurement ensembles $\A$. Figure \ref{fig:MeasurementTest} shows the average relative approximation error when reconstructing $100$ randomly drawn $\X_\star$ for varying $m$. We compare here three types of measurements satisfying the assumption of Theorems \ref{thm:SmallBallRIP} and \ref{GaussianRIP}: first, operators $\A$ whose components $\AA_i$ have i.i.d.\ standard normal entries. Second, operators $\A$ whose components $\AA_i$ have i.i.d.\ log-normal distributed entries. Third, operators $\A$ whose components satisfy $\AA_i = \a_i \a_i^\top$ for Gaussian random vectors $\a_i$, $i \in [m]$. While the first two choices (Figure \ref{fig:MeasurementTestA}) allow to re-use the setting of Section \ref{sec:Numerics1}, i.e., $\X_\star \in K_{20,20}^1 \in \R^{20 \times 300}$, the Gaussian rank-$1$ measurements (Figure \ref{fig:MeasurementTestB}) require square matrices. We thus consider $\X_\star \in K_{10,10}^1 \subset \R^{50 \times 50}$ in the third case. In all cases, the noise level is $\| \Eta \|_{F} = 0.1 \| \X_\star \|_{F}$. \\
As Figure \ref{fig:MeasurementTest} shows, Algorithm \ref{alg:AM} yields good approximation for all three types of measurements. More important, the error is already close to noise level for a number of measurements far below the number required in mere sparse/low-rank approximation. To substantiate this point, we provide for comparison the outcome when using AltMinSense, a state-of-the-art method only using low-rankness \cite{jain2013low}.  

\subsection{Algorithm \ref{alg:AM} vs Algorithm \ref{alg:PALM}}
\label{sec:NumericsAlg1vsAlg3}

\rev{The third experiment illustrates the difference in performance between Algorithm \ref{alg:AM} and Algorithm \ref{alg:PALM}. We recover a randomly drawn $\X_\star \in K_{20,20}^1 \subset \R^{20\times 300}$ from $m = 240$ Gaussian measurements with additive noise of $\| \Eta \|_2 = 0.1 \| \X_\star \|_F$. Both algorithms run with the same parameters $\abg$ and are initialized by the randomly perturbed ground-truth, see Section \ref{Numerics}. Algorithm \ref{alg:AM} uses $20$ outer iterations with stopping criterion $\max\{ \| \U_{k+1}-\U_k \|, \| \V_{k+1} - \V_k \| \} \le 10^{-2}$ and $10^{1}, 10^{2}, 10^{3}$ inner iterations of Algorithm \ref{alg:ProxDesc} with stopping criterion $\| \x_{k+1} - \x_k \|_2 \le 10^{-6}$. Algorithm \ref{alg:PALM} uses $2 \cdot 10^{4}$ iterations with stopping criterion $\max\{ \| \U_{k+1}-\U_k \|, \| \V_{k+1} - \V_k \| \} \le 10^{-7}$.

Figure \ref{fig:Runtime} illustrates that in running time Algorithm \ref{alg:PALM} clearly outperforms Algorithm \ref{alg:AM} independent of the number of iterations one uses in the inner iterations in Algorithm \ref{alg:ProxDesc}. Nevertheless, the additional step-size parameters $\lambda_k$ and $\mu_k$ apparently have a strong influence on the reconstruction performance. Another interesting observation is that the inner iteration number of Algorithm \ref{alg:AM} seems to exhibit a sharp transition between a regime where there are too few inner iterations per alternating step (solid green line) and a regime where the number of inner iterations suffices (solid red and solid blue line). Increasing the number of iterations beyond this point doesn't have much impact on the reconstruction accuracy as the solid red and solid blue line show.
}

\subsection{Influence of initialization quality}
\label{sec:Init}

\begin{figure}[ht!]
 	\centering
 	\captionsetup{width=0.95\linewidth}
 	\begin{subfigure}[b]{0.45\textwidth}
 		\includegraphics[width=\textwidth]{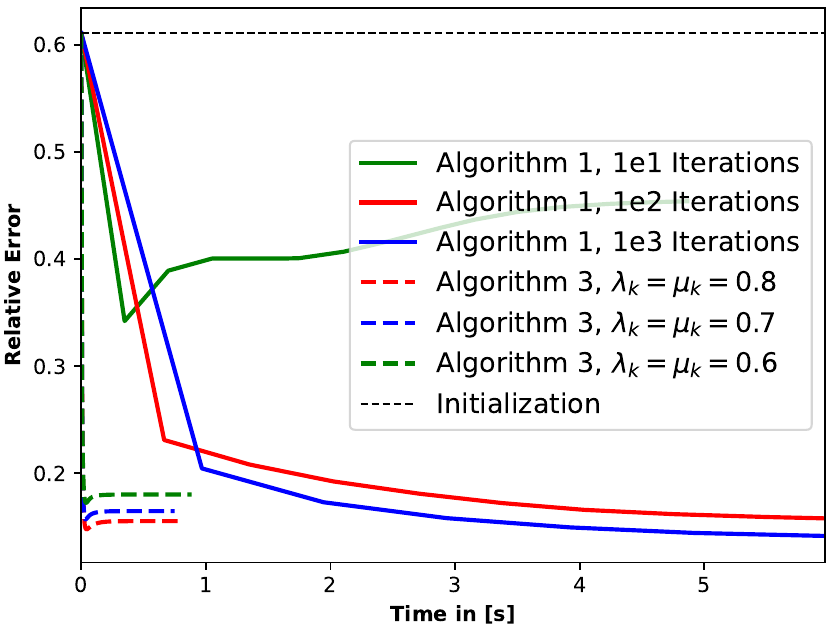}
 		\subcaption{Runtime comparison.}
 		\label{fig:Runtime}
 	\end{subfigure}
 	\quad
 	\begin{subfigure}[b]{0.45\textwidth}
 		\includegraphics[width=\textwidth]{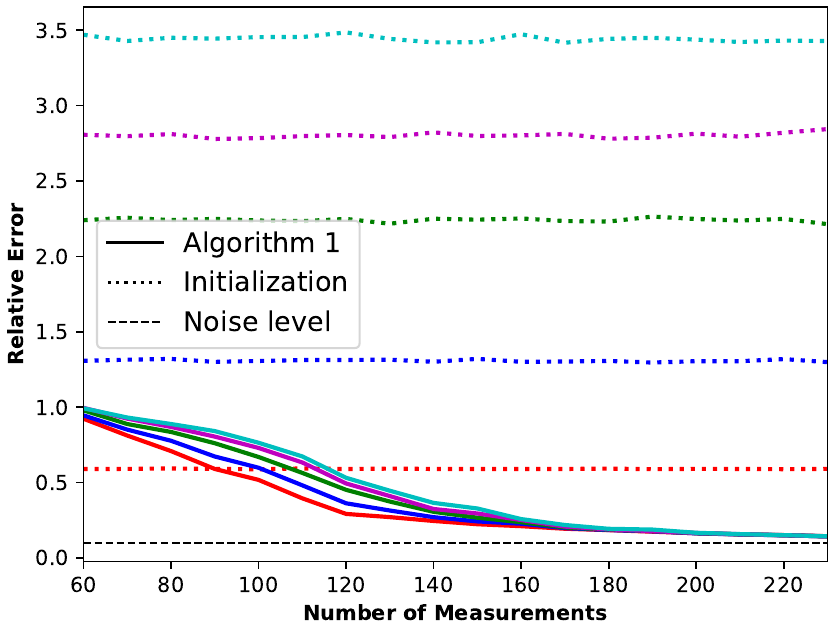}
 		\subcaption{Initialization comparison.}
 		\label{fig:Initialization}
 	\end{subfigure}
  \caption{Runtime comparison for Algorithms \ref{alg:AM} and \ref{alg:PALM}, and comparison of initialization quality. The approximation error is measured relative to $\Vert \X_\star \Vert_F$ in both cases. \\ 
  In \ref{fig:Initialization}, lines of the same color belong to the same initialization set-up. The noise level $\| \Eta \|_2$ is provided as a benchmark for the reconstruction accuracy.}
\label{fig:RuntimeAndInit}
\end{figure}

\rev{In a fourth experiment, we re-use the experimental set-up from Section \ref{sec:Numerics2} with Gaussian $\A$ but vary the quality of initialization. To this end, we increase the random perturbation added to the ground-truth SD $\U_\star$ and $\V_\star$ step by step from a relative strength of $0.4$ to $1.6$, cf.\ beginning of Section \ref{Numerics}.

Figure \ref{fig:Initialization} illustrates that the initialization quality only has a strong impact on the reconstruction accuracy if $m$ is close to the information theoretic lower bound $\rank(\X_\star)(s_1+s_2) = 40$. As soon as the oversampling factor is sufficiently large, the convergence radius of Algorithm \ref{alg:AM} appears to grow considerably. Note that the information theoretic lower bound for guaranteed reconstruction would be $m~\ge~\rank(\X_\star) (n_1 + n_2) = 320$ if only using low-rank structure and $m \ge s_1 s_2 = 400$ if only using sparsity structure.
}

\subsection{A comparison with SPF} \label{Numerics3}

\begin{figure}[ht!]
 	\centering
 	\captionsetup{width=0.95\linewidth}
 	\begin{subfigure}[b]{0.45\textwidth}
 		\includegraphics[width=\textwidth]{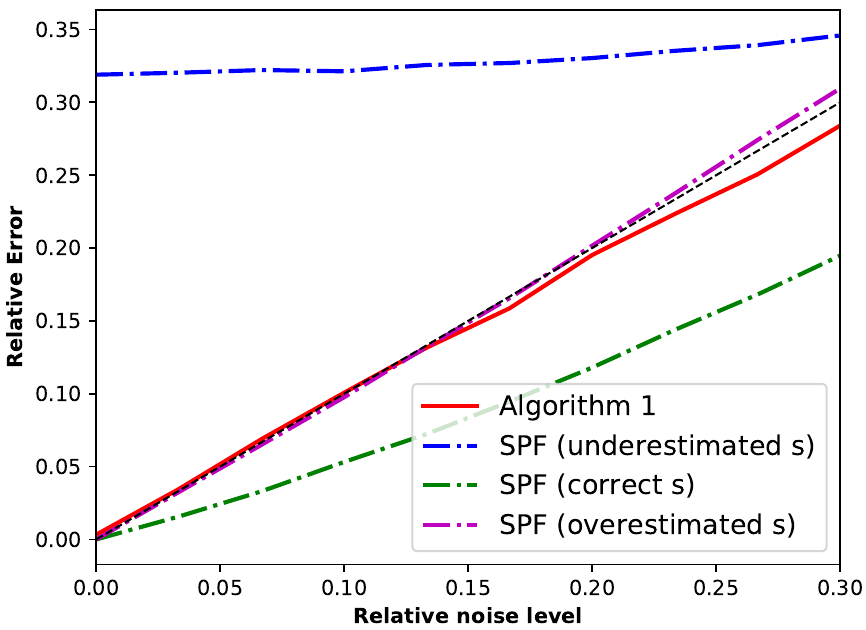}
 		\subcaption{Jointly sparse $\X_\star$.}
 		\label{fig:NoiseTestA}
 	\end{subfigure}
 	\quad
 	\begin{subfigure}[b]{0.45\textwidth}
 		\includegraphics[width=\textwidth]{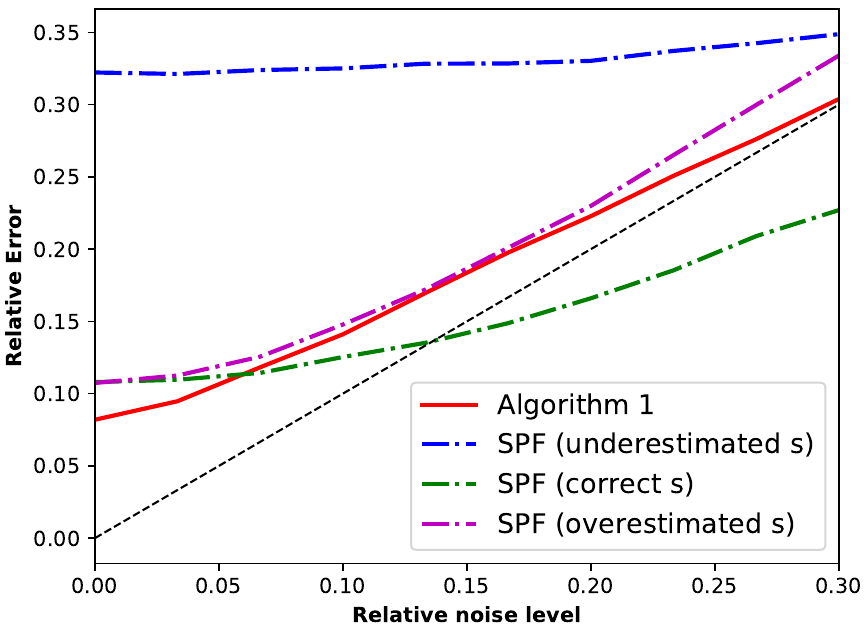}
 		\subcaption{Effectively sparse $\X_\star$.}
 		\label{fig:NoiseTestB}
 	\end{subfigure}
  \caption{Comparison of Algorithm \ref{alg:AM} and SPF \cite{lee2013near} for various noise levels. The approximation error and the noise level are measured relative to $\Vert \X_\star \Vert_F$. For convenience, the identity map is given in form of a black dashed line.}
\label{fig:NoiseTest}
\end{figure}

After having provided empirical evidence for our theoretical results, we now turn to the comparison of our method with its state-of-the-art counterpart SPF  \cite{lee2013near}. To the best of our knowledge, SPF is the only available algorithm so far, which simultaneously leverages low-rankness and sparsity constraints and comes with near-optimal recovery guarantees (not relying on a special structure of $\A$ as in \cite{bahmani2016near}). As \cite{lee2013near} contains exhaustive numerical comparisons of SPF and low-rank/sparse reconstruction strategies based on convex relaxation, SPF suffices for numerical benchmark tests of our method. 

\rev{

\paragraph{} In our first experiment, we use toy data and create rank-$1$ ground-truths in $\X_\star \in \R^{20\times 300}$ that are (i) jointly sparse, i.e., $\X_\star \in \bar{S}_{20,20}^1$, by randomly drawing a support set and filling it with random Gaussian entries, and (ii) effectively sparse as described in the beginning of Section \ref{Numerics}, i.e., $\X_\star \in K_{20,20}^1$. In Figure \ref{fig:NoiseTest} we compare the performance of SPF and Algorithm \ref{alg:AM} for Gaussian measurements and different noise levels $\| \Eta \|_2$. Whereas the parameters $\abg$ have been tuned as described in the beginning of Section \ref{Numerics}, we compare three versions of SPF: first, with correct sparsity parameter $s = 20$, second, with underestimated sparsity parameter $s = 10$, and, third, with overestimated sparsity parameter $s = 40$. The experiments are averaged over $100$ random draws of $\X_\star$.\\
In case (i), Figure \ref{fig:NoiseTestA} shows that if the sparsity parameter of SPF is exactly tuned, it easily outperforms Algorithm \ref{alg:AM}. This is not surprising since SPF exploits the joint support structure of $\X_\star$. Two observations are, however, notable. First, the performance of SPF strongly deteriorates if its sparsity parameter $s$ is chosen slightly differently. Second, contrary to what the additive term $\sqrt{\delta}$ in the error bound of Theorem \ref{ApproxX} suggests, Algorithm \ref{alg:AM} works well for arbitrary low noise levels. This suggests that our analysis for $\bar{S}_{s_1,s_2}^R$ is too pessimistic and can be improved.\\
In case (ii), Figure \ref{fig:NoiseTestB} shows that SPF is notably hampered by the non-exact sparsity of the ground-truth if the noise level is small. Algorithm \ref{alg:AM} appears here to be more stable. Moreover, the fact that both algorithms fail in recovering $\X_\star$ for $\Eta \to \0$ match our analysis of the general setting in Theorem \ref{ApproxX}, which includes a error saturation depending on the injectivity constant $\delta$.
}

\begin{figure}[ht!]
	\centering
	\captionsetup{width=0.95\linewidth}
	\begin{subfigure}[b]{0.45\textwidth}
		\includegraphics[width=\textwidth]{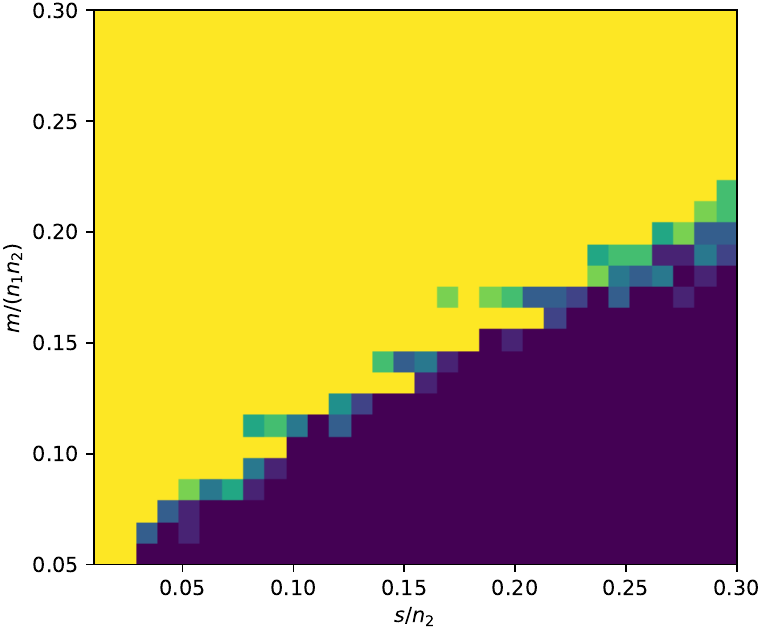}
		\caption{SPF, Best Approximation}
		\label{fig:a}
	\end{subfigure}
	\quad 
	\begin{subfigure}[b]{0.45\textwidth}
		\includegraphics[width=\textwidth]{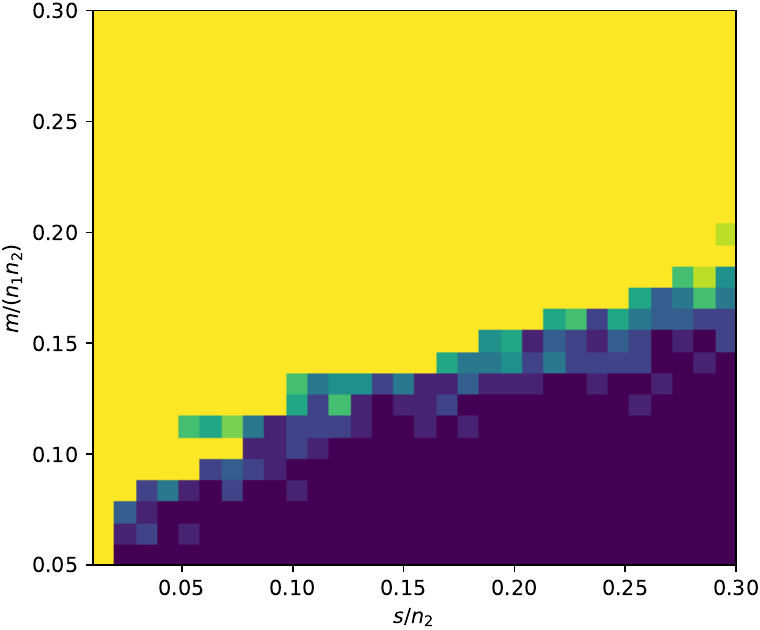}
		\caption{Algorithm \ref{alg:AM}, Best Approximation}
		\label{fig:b}
	\end{subfigure}\\
    \begin{subfigure}[b]{0.45\textwidth}
    	\includegraphics[width=\textwidth]{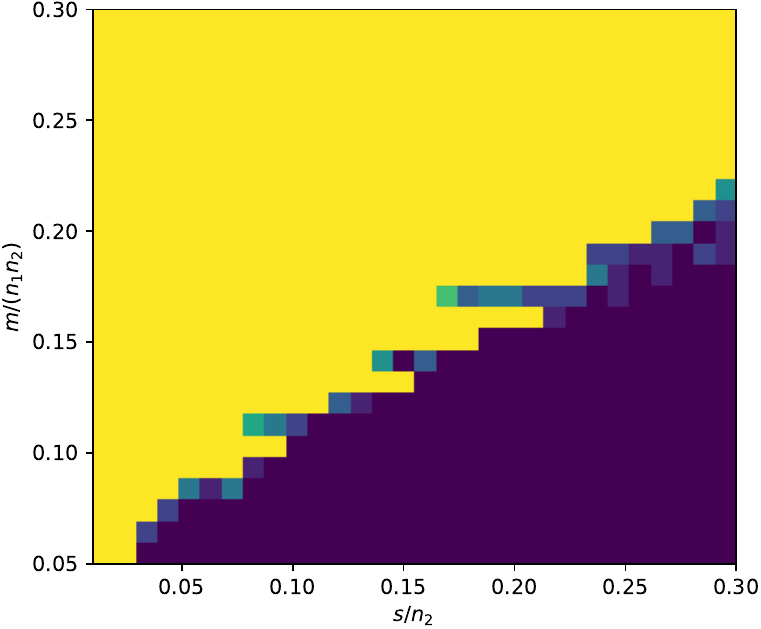}
    	\caption{SPF, Discrepancy Principle}
    	\label{fig:c}
    \end{subfigure}
    \quad
	\begin{subfigure}[b]{0.45\textwidth}
		\includegraphics[width=\textwidth]{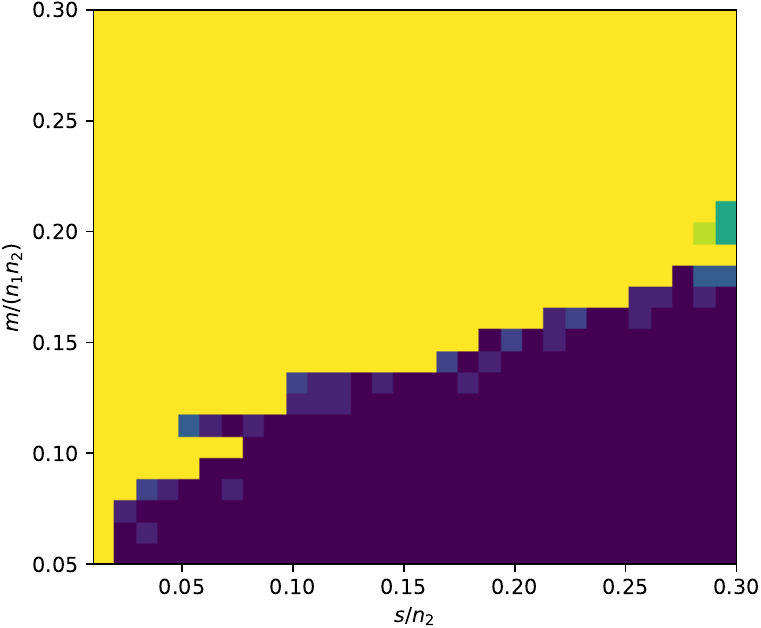}
		\caption{Algorithm \ref{alg:AM}, Discrepancy Principle}
		\label{fig:d}
	\end{subfigure}
	\caption{Phase transition diagrams comparing SPF and Algorithm \ref{alg:AM}. Empirical recovery probability is depicted by color from zero (blue) to one (yellow).}
	\label{fig:Grayscale}
\end{figure}

\paragraph{} In the second experiment depicted in Figure \ref{fig:Grayscale} we compare for $s/n_2 \in [0,0.3]$ and $m/(n_1 n_2) \in [0.05,0.3]$ the number of successful recoveries of $10$ randomly drawn $\X_\star \in K_{16,s}^3 \subset \R^{16\times 100}$ from $m$ Gaussian measurements. We set the noise level to $\norm{\Eta}{F} = 0.2 \Vert \X_\star \Vert_F$ and count the reconstruction successful if $\Vert \X_\star - \X_\text{SPF} \Vert_F/\Vert \X_\star \Vert_F \le 0.4$ resp.\ $\Vert \X_\star - \X_\abg \Vert_F/\Vert \X_\star \Vert_F \le 0.4$. The sparsity parameter $s'$ of SPF is optimized over the grid $\curly{5,10,\dots,45,50}$. In Figures \ref{fig:Grayscale} (a)-(b) we compare the two algorithms if their respective parameters are tuned under knowledge of $\X_\star$ (Best Approximation), whereas in Figures \ref{fig:Grayscale} (c)-(d) only the noise level $\norm{\Eta}{F}$ is known (Discrepancy Principle). In the latter case, the parameter $s'$ of SPF minimizes $| \| \X_\star - \X_{\text{SPF}} \|_{F} - \| \Eta \|_F |$ and the heuristic for Algorithm \ref{alg:AM} reduces $\mu$ until $| \| \X_\star - \X_\abg \|_{F} - \| \Eta \|_F |$ stops decreasing. Both algorithms show a comparable performance. Nevertheless, the theoretical guarantees of Algorithm \ref{alg:AM} cover a considerably larger class of signals and output non-orthogonal decompositions.

\begin{figure}[!htb]
 	\centering
 	\captionsetup{width=0.9\linewidth}
 	\begin{subfigure}[b]{0.15\textwidth}
 		\includegraphics[width=\textwidth]{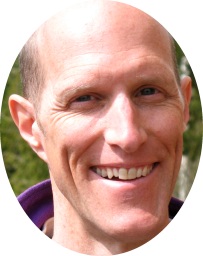}
 	\end{subfigure}
 	\quad
 	\begin{subfigure}[b]{0.15\textwidth}
 		\includegraphics[width=\textwidth]{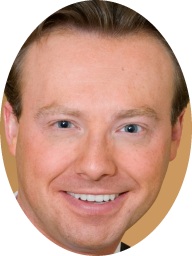}
 	\end{subfigure}
 	\quad
 	\begin{subfigure}[b]{0.15\textwidth}
 		\includegraphics[width=\textwidth]{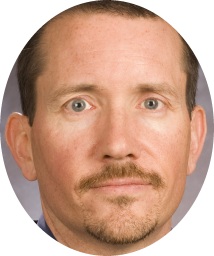}
 	\end{subfigure}
 	\quad
 	\begin{subfigure}[b]{0.15\textwidth}
 		\includegraphics[width=\textwidth]{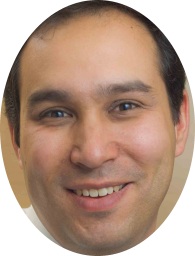}
 	\end{subfigure}
 	\quad
 	\begin{subfigure}[b]{0.15\textwidth}
 		\includegraphics[width=\textwidth]{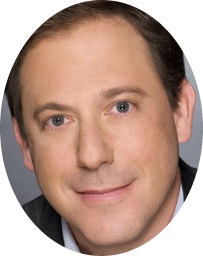}
 	\end{subfigure}
 	\quad
  \caption{ \rev{Five samples from original face data.} }
\label{fig:OriginalFaces}
\end{figure}

\rev{

\paragraph{} In a third experiment, we repeat a simulation from \cite[Section 8.3]{fornasier2018robust} on real data. We choose ten faces from the ``10k US Adult Faces Data Base'' \cite{bainbridge2013intrinsic}, cf.\ Figure \ref{fig:OriginalFaces}, reduce their resolution to 64x44 pixels and their color range to gray-scale, and apply a multi-layer wavelet transform with Haar-wavelets to obtain effectively sparse representations. The resulting ten $2944$-dimensional coefficient vectors are then used to build a ground truth matrix $\X_\star \in \R^{10\times 2944}$ which is re-scaled to unit Frobenius norm. Due to the use of Haar-wavelets this matrix is not very close to sparse but has an effective sparsity level of about $s = 70$ which we measure as described in the beginning of Section \ref{Numerics}. Though not rank-deficient, $\X_\star$ has effective rank $\boldsymbol{r}(\X_\star) \approx 1.5$ where\begin{align*}
    \boldsymbol{r}(\X_\star) = \frac{\| \X_\star \|_*}{\| \X_\star \|_{2\to 2}} \in [ 1,\rank(\X_\star) ]
\end{align*}
is the ratio of nuclear and operator norm and serves as a relaxed measure of low-rankness of a matrix similar to effective sparsity for vectors. In particular, $\X_\star$ is well approximated by low-rank matrices. As in \cite{fornasier2018robust}, $\mathcal{A}$ is a Gaussian operator and the noise level is set to $\| \Eta \|_2 = 0.2 \| \X_\star \|_F$. We set $m$ as ten times the information theoretic limit $R(n_1+s)$ where $R$ is the rank parameter used for SPF and Algorithm \ref{alg:PALM} in the experiments below (since the computational load is much higher in this experiment due to the largely increased ambient dimension, we do not use the less efficient Algorithm \ref{alg:AM} here). We set the additional step-size parameters of Algorithm \ref{alg:PALM} to $\lambda_k = \mu_k = 1$ and initialize both algorithms by spectral initialization, see Section \ref{Numerics}.\\
\begin{table}[!htb]
    \centering
    \blue{
    \begin{tabular}{ || m{3cm} || m{1.6cm}|m{1.6cm}|m{1.6cm} || } 
        \hline
        Setting & $R = 2$ & $R = 3$ & $R = 4$ \\
        \hline \hline
        SPF (B.A.) & 0.2927 & 0.2628 & 0.2507    \\
        Algorithm \ref{alg:PALM} (B.A.) & \textbf{0.2647} & \textbf{0.2471} & \textbf{0.2452} \\
        \hline
        SPF (D.P.) & 0.3017 & 0.2905 & 0.2645 \\
        Algorithm \ref{alg:PALM} (D.P.) & \textbf{0.2647} & \textbf{0.2502} & \textbf{0.2482} \\
        \hline
        $\| \X_\star - (\X_\star)_R \|_F$ & 0.1341 & 0.1130 & 0.0968 \\
        \hline
    \end{tabular}
    \caption{\blue{Comparison of SPF and Algorithm \ref{alg:PALM} for different choices of $R$ (the error produced by the best rank-$R$ term approximation $(\X_\star)_R$ of $\X_\star$ is given as a benchmark). Depicted is the relative approximation error measured in Frobenius norm. The parameters $s$ (for SPF) resp.\ $\abg$ (for Algorithm \ref{alg:PALM}) have been tuned on a discrete search grid either using \textbf{B}est \textbf{A}pproximation, i.e., minimizing $\| X_{\text{rec}} - \X_\star \|_F$, or \textbf{D}iscrepancy \textbf{P}rinciple, i.e., minimizing $| \| \mathcal{A}(X_{\text{rec}}) - y \|_2 - \| \eta \|_2|$.}} \label{fig:Table1}
    }
\end{table}

In Table \ref{fig:Table1} we compare the full matrix reconstruction performance of SPF and Algorithm \ref{alg:PALM} for different choices of the rank hyper-parameter $R$. As a benchmark, the error produced by best rank-$R$ term approximation is reported as well; this is the best achievable error without added noise and under full knowledge of $\X_\star$. The results show that Algorithm \ref{alg:PALM} does well in comparison with SPF. As also observed in \cite{fornasier2018robust}, the performance of SPF stronger deteriorates than the one of Algorithm \ref{alg:PALM} if the hyper-parameters are tuned under exclusive knowledge of the noise-level $\| \Eta \|_2$.

Finally, we provide a visual comparison of the reconstruction results. To this end, we set the noise level to zero, increase the oversampling factor from ten to twenty, and set $R = 5$. Figure \ref{fig:Face1} illustrates that SPF and Algorithm \ref{alg:PALM} produce qualitatively different approximations despite the comparable $\ell_2$-reconstruction errors. As already mentioned in \cite{fornasier2018robust}, SPF tends to oversimplify the image by producing large pixel areas of uniform gray-level. Comparing Figure \ref{fig:Face1}(c) to \cite[Figure 8(c)]{fornasier2018robust} further shows that Algorithm \ref{alg:PALM} and ATLAS in \cite{fornasier2018robust} lead to similar results. This suggests that the modifications to ATLAS which we proposed in this paper to improve the theoretical analysis do not hamper its empirical performance.

\begin{figure}[tb]
 	\centering
 	\captionsetup{width=0.9\linewidth}
 	\begin{subfigure}[b]{0.2\textwidth}
 		\includegraphics[width=\textwidth]{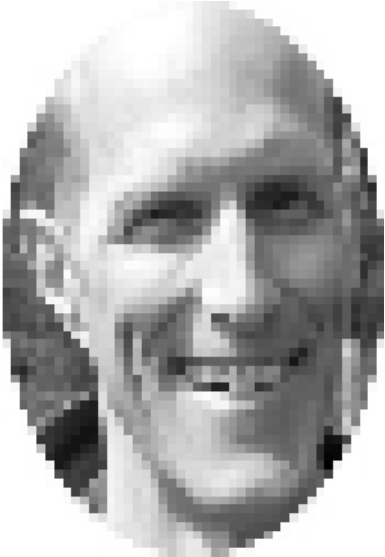}
 		\subcaption{Ground truth}
 	\end{subfigure}
 	\qquad
 	\begin{subfigure}[b]{0.2\textwidth}
 		\includegraphics[width=\textwidth]{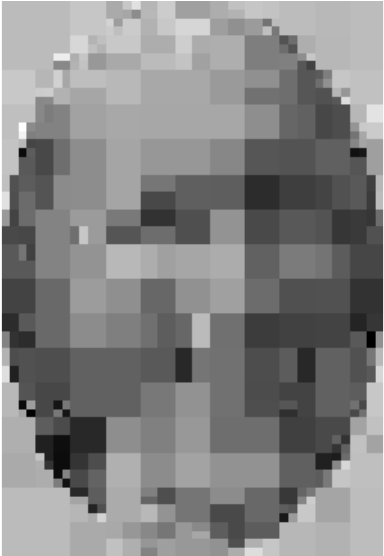}
 		\subcaption{SPF}
 	\end{subfigure}
 	\qquad
 	\begin{subfigure}[b]{0.2\textwidth}
 		\includegraphics[width=\textwidth]{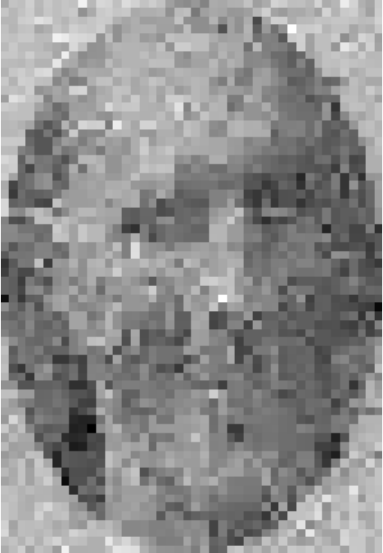}
 		\subcaption{Algorithm \ref{alg:PALM}}
 	\end{subfigure}
 	\qquad
 	\begin{subfigure}[b]{0.2\textwidth}
 		\includegraphics[width=\textwidth]{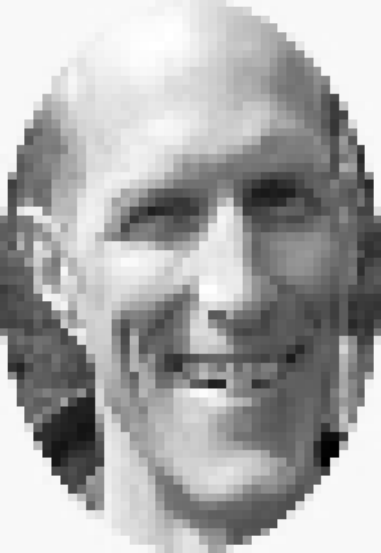}
 		\subcaption{Rank-$R$ SVD}
 	\end{subfigure}
  \caption{ \blue{Comparison of the first face and its reconstructions.} }
\label{fig:Face1}
\end{figure}

}


\section{Discussion and Open Questions} \label{Conclusion}

In this paper we proposed a multi-penalty approach to recover low-rank matrices with sparsity structure from incomplete and inaccurate linear measurements. To improve on the results of \cite{fornasier2018robust}, we introduced a conceptually different functional $J_\abg^R$ and a revised signal set $K_{s_1,s_2}^R$, the combination of which allowed us a clean analysis of the approximation quality of global minimizers. In particular, the new approach encompasses heavy-tailed measurement ensembles and structured rank-1 measurements. \rev{In the special case of subgaussian measurements, our results reach up to log-factors the same sample complexity as existing theoretical guarantees for SPF that are known to be near-optimal, cf.\ Remark \ref{rem:dim}. At the same time, our method tackles the recovery of a significantly larger class of matrices, namely matrices with non-orthogonal rank-$1$ decompositions and effectively sparse components and the analysis of $J_\abg^R$ is far more elementary than the one of SPF.

Several intriguing questions remain open at this point. First, the shape of our error bounds appears to be suboptimal for the particular case of jointly sparse ground-truths in $\bar{S}_{s_1,s_2}^R$ to which the analysis in \cite{lee2013near} is restricted, see Theorem \ref{ApproxX} and the consequent discussion. As mentioned above, this is partly due to the fact that minimizers of $J_\abg^R$ in general consist of non-orthogonal matrices. If one could show (near) orthogonality of such minimizers under the assumption $\X_\star \in \bar{S}_{s_1,s_2}^R$, then it would be possible to derive a scaling invariant signal set of similar complexity with scaling invariant injectivity property and to remove the additive $\sqrt{\delta}$-term from \eqref{Approximation}. }

Second, we didn't solve the problem of initialization here, cf.\ Remark \ref{rem:Haeffele}. A crucial task for the near future is thus to provide an initialization procedure that guarantees computation of global minimizers of $J_\abg^R$ via Algorithms \ref{alg:AM} \& \ref{alg:PALM}. Spectral initialization, the state-of-the-art procedure for non-convex methods in low-rank matrix sensing, cf.\ \cite{jain2013low}, certainly works for $m$ sufficiently large. However, it is unlikely that spectral initialization can be used to reconstruct at the information theoretic limit if sparsity and low-rankness are considered simultaneously.

Finally, further structured measurement ensembles should be examined like, e.g., sub-sampled circulant matrices. This kind of measurements naturally appears in applications like blind deconvolution \cite{haykin1994blind}.


%


\section*{Acknowledgments}

The author gratefully acknowledges funding by the Deutsche Forschungsgemeinschaft (DFG, German Research Foundation) through the project CoCoMIMO funded within the priority program SPP 1798 {\it{Compressed Sensing in Information Processing}} (COSIP). The author furthermore thanks the anonymous reviewers for their detailed and helpful comments.

\bibliographystyle{IEEEtranS}
\bibliography{mybibM,CSBib}

\begin{thebibliography}{10}
\providecommand{\url}[1]{#1}
\csname url@samestyle\endcsname
\providecommand{\newblock}{\relax}
\providecommand{\bibinfo}[2]{#2}
\providecommand{\BIBentrySTDinterwordspacing}{\spaceskip=0pt\relax}
\providecommand{\BIBentryALTinterwordstretchfactor}{4}
\providecommand{\BIBentryALTinterwordspacing}{\spaceskip=\fontdimen2\font plus
\BIBentryALTinterwordstretchfactor\fontdimen3\font minus
  \fontdimen4\font\relax}
\providecommand{\BIBforeignlanguage}[2]{{%
\expandafter\ifx\csname l@#1\endcsname\relax
\typeout{** WARNING: IEEEtranS.bst: No hyphenation pattern has been}%
\typeout{** loaded for the language `#1'. Using the pattern for}%
\typeout{** the default language instead.}%
\else
\language=\csname l@#1\endcsname
\fi
#2}}
\providecommand{\BIBdecl}{\relax}
\BIBdecl

\bibitem{attouch2010proximal}
H.~Attouch, J.~Bolte, P.~Redont, and A.~Soubeyran, ``Proximal alternating
  minimization and projection methods for nonconvex problems: An approach based
  on the kurdyka-{\l}ojasiewicz inequality,'' \emph{Mathematics of Operations
  Research}, vol.~35, no.~2, pp. 438--457, 2010.

\bibitem{bahmani2016near}
S.~Bahmani and J.~Romberg, ``Near-optimal estimation of simultaneously sparse
  and low-rank matrices from nested linear measurements,'' \emph{Information
  and Inference: A Journal of the IMA}, vol.~5, no.~3, pp. 331--351, 2016.

\bibitem{bainbridge2013intrinsic}
W.~A. Bainbridge, P.~Isola, and A.~Oliva, ``The intrinsic memorability of face
  photographs.'' \emph{Journal of Experimental Psychology: General}, vol. 142,
  no.~4, p. 1323, 2013.

\bibitem{BLUMENSATH2009265}
T.~Blumensath and M.~E. Davies, ``Iterative hard thresholding for compressed
  sensing,'' \emph{Applied and Computational Harmonic Analysis}, vol.~27,
  no.~3, pp. 265 -- 274, 2009.

\bibitem{bolte2007lojasiewicz}
J.~Bolte, A.~Daniilidis, and A.~Lewis, ``The {\l}ojasiewicz inequality for
  nonsmooth subanalytic functions with applications to subgradient dynamical
  systems,'' \emph{SIAM Journal on Optimization}, vol.~17, no.~4, pp.
  1205--1223, 2007.

\bibitem{bolte2014proximal}
J.~Bolte, S.~Sabach, and M.~Teboulle, ``Proximal alternating linearized
  minimization for nonconvex and nonsmooth problems,'' \emph{Mathematical
  Programming}, vol. 146, no. 1-2, pp. 459--494, 2014.

\bibitem{candes2011tight}
E.~J. Candes and Y.~Plan, ``Tight oracle inequalities for low-rank matrix
  recovery from a minimal number of noisy random measurements,'' \emph{IEEE
  Transactions on Information Theory}, vol.~57, no.~4, pp. 2342--2359, 2011.

\bibitem{d2005direct}
A.~d'Aspremont, L.~E. Ghaoui, M.~I. Jordan, and G.~R. Lanckriet, ``A direct
  formulation for sparse {PCA} using semidefinite programming,'' in
  \emph{Advances in neural information processing systems}, 2005, pp. 41--48.

\bibitem{de2012decoupling}
V.~De~la Pena and E.~Gin{\'e}, \emph{Decoupling: from dependence to
  independence}.\hskip 1em plus 0.5em minus 0.4em\relax Springer Science \&
  Business Media, 2012.

\bibitem{eisenmann2021riemannian}
H.~Eisenmann, F.~Krahmer, M.~Pfeffer, and A.~Uschmajew, ``Riemannian
  thresholding methods for row-sparse and low-rank matrix recovery,''
  \emph{ArXiv:2103.02356}, 2021.

\bibitem{fernique1975regularite}
X.~Fernique, ``Regularit{\'e} des trajectoires des fonctions al{\'e}atoires
  gaussiennes,'' in \emph{Ecole d’Et{\'e} de Probabilit{\'e}s de Saint-Flour
  IV—1974}.\hskip 1em plus 0.5em minus 0.4em\relax Springer, 1975, pp. 1--96.

\bibitem{fornasier2018robust}
M.~Fornasier, J.~Maly, and V.~Naumova, ``Robust recovery of low-rank matrices
  with non-orthogonal sparse decomposition from incomplete measurements,''
  \emph{Applied Mathematics and Computation}, vol. 392, 2021.

\bibitem{foucart2019jointly}
S.~Foucart, R.~Gribonval, L.~Jacques, and H.~Rauhut, ``Jointly low-rank and
  bisparse recovery: Questions and partial answers,'' \emph{Analysis and
  Applications}, vol.~18, no.~01, pp. 25--48, 2020.

\bibitem{foucart:2013}
S.~Foucart and H.~Rauhut, \emph{A Mathematical Introduction to Compressive
  Sensing}.\hskip 1em plus 0.5em minus 0.4em\relax Birkh\"auser Basel, 2013.

\bibitem{geppert2019sparse}
J.~Geppert, F.~Krahmer, and D.~St{\"o}ger, ``Sparse power factorization:
  balancing peakiness and sample complexity,'' \emph{Advances in Computational
  Mathematics}, vol.~45, no.~3, pp. 1711--1728, 2019.

\bibitem{naumova2016}
M.~Grasmair and V.~Naumova, ``Conditions on optimal support recovery in
  unmixing problems by means of multi-penalty regularization,'' \emph{Inverse
  Problems}, vol.~32, no.~10, p. 104007, 2016.

\bibitem{haeffele2019structured}
B.~D. Haeffele and R.~Vidal, ``Structured low-rank matrix factorization: Global
  optimality, algorithms, and applications,'' \emph{IEEE Transactions on
  Pattern Analysis and Machine Intelligence}, vol.~42, no.~6, pp. 1468--1482,
  2019.

\bibitem{haykin1994blind}
S.~Haykin, ``The blind deconvolution problem,'' \emph{Blind Deconvolution},
  p.~1, 1994.

\bibitem{iwen2014robust}
M.~Iwen, A.~Viswanathan, and Y.~Wang, ``Robust sparse phase retrieval made
  easy,'' \emph{Applied and Computational Harmonic Analysis}, vol.~42, no.~1,
  pp. 135--142, 2017.

\bibitem{jain2013low}
P.~Jain, P.~Netrapalli, and S.~Sanghavi, ``Low-rank matrix completion using
  alternating minimization,'' \emph{Proceedings of the forty-fifth Annual ACM
  Symposium on Theory of Computing}, pp. 665--674, 2013.

\bibitem{klibanov1995phase}
M.~V. Klibanov, P.~E. Sacks, and A.~V. Tikhonravov, ``The phase retrieval
  problem,'' \emph{Inverse problems}, vol.~11, no.~1, p.~1, 1995.

\bibitem{kliesch2019simultaneous}
M.~Kliesch, S.~J. Szarek, and P.~Jung, ``Simultaneous structures in convex
  signal recovery—revisiting the convex combination of norms,''
  \emph{Frontiers in Applied Mathematics and Statistics}, vol.~5, p.~23, 2019.

\bibitem{kueng2017low}
R.~Kueng, H.~Rauhut, and U.~Terstiege, ``Low rank matrix recovery from rank one
  measurements,'' \emph{Applied and Computational Harmonic Analysis}, vol.~42,
  no.~1, pp. 88--116, 2017.

\bibitem{kurdyka1998gradients}
K.~Kurdyka, ``On gradients of functions definable in o-minimal structures,'' in
  \emph{Annales de l'institut Fourier}, vol.~48, no.~3, 1998, pp. 769--783.

\bibitem{lee2016blind}
K.~Lee, Y.~Li, M.~Junge, and Y.~Bresler, ``Blind recovery of sparse signals
  from subsampled convolution,'' \emph{IEEE Transactions on Information
  Theory}, vol.~63, no.~2, pp. 802--821, 2016.

\bibitem{lee2013near}
K.~Lee, Y.~Wu, and Y.~Bresler, ``Near-optimal compressed sensing of a class of
  sparse low-rank matrices via sparse power factorization,'' \emph{IEEE
  Transactions on Information Theory}, vol.~64, no.~3, pp. 1666--1698, 2018.

\bibitem{li2013global}
G.~Li, ``Global error bounds for piecewise convex polynomials,''
  \emph{Mathematical Programming}, vol. 137, no. 1-2, pp. 37--64, 2013.

\bibitem{lojasiewicz1963propriete}
S.~Lojasiewicz, ``Une propri{\'e}t{\'e} topologique des sous-ensembles
  analytiques r{\'e}els,'' \emph{Les {\'e}quations aux d{\'e}riv{\'e}es
  partielles}, vol. 117, pp. 87--89, 1963.

\bibitem{magdon2017np}
M.~Magdon-Ismail, ``Np-hardness and inapproximability of sparse pca,''
  \emph{Information Processing Letters}, vol. 126, pp. 35--38, 2017.

\bibitem{mendelson2014learning}
S.~Mendelson, ``Learning without concentration,'' in \emph{Conference on
  Learning Theory}, 2014, pp. 25--39.

\bibitem{mordukhovich2006variational}
B.~S. Mordukhovich, \emph{Variational analysis and generalized differentiation
  I: Basic theory}.\hskip 1em plus 0.5em minus 0.4em\relax Springer Science \&
  Business Media, 2006, vol. 330.

\bibitem{naumova2014minimization}
V.~Naumova and S.~Peter, ``Minimization of multi-penalty functionals by
  alternating iterative thresholding and optimal parameter choices,''
  \emph{Inverse Problems}, vol.~30, no.~12, p. 125003, 2014.

\bibitem{oymak2015simultaneously}
S.~Oymak, A.~Jalali, M.~Fazel, Y.~C. Eldar, and B.~Hassibi, ``Simultaneously
  structured models with application to sparse and low-rank matrices,''
  \emph{IEEE Transactions on Information Theory}, vol.~61, no.~5, pp.
  2886--2908, 2015.

\bibitem{Plan2013LP}
Y.~Plan and R.~Vershynin, ``One-bit compressed sensing by linear programming,''
  \emph{Communications on Pure and Applied Mathematics}, vol.~66, no.~8, pp.
  1275--1297, 2013.

\bibitem{Plan2014}
------, ``Dimension reduction by random hyperplane tessellations,''
  \emph{Discrete {\&} Computational Geometry}, vol.~51, no.~2, pp. 438--461,
  2014.

\bibitem{recht2010}
B.~Recht, M.~Fazel, and P.~A. Parrilo, ``Guaranteed minimum-rank solutions of
  linear matrix equations via nuclear norm minimization,'' \emph{SIAM Review},
  vol.~52, no.~3, pp. 471--501, 2010.

\bibitem{rockafellar2009variational}
R.~T. Rockafellar and R.~J.-B. Wets, \emph{Variational analysis}.\hskip 1em
  plus 0.5em minus 0.4em\relax Springer Science \& Business Media, 2009, vol.
  317.

\bibitem{talagrand2006generic}
M.~Talagrand, \emph{The generic chaining: upper and lower bounds of stochastic
  processes}.\hskip 1em plus 0.5em minus 0.4em\relax Springer Science \&
  Business Media, 2006.

\bibitem{talagrand2014upper}
------, \emph{Upper and lower bounds for stochastic processes: modern methods
  and classical problems}.\hskip 1em plus 0.5em minus 0.4em\relax Springer
  Science \& Business Media, 2014, vol.~60.

\bibitem{tropp2015convex}
J.~A. Tropp, ``Convex recovery of a structured signal from independent random
  linear measurements,'' in \emph{Sampling Theory, a Renaissance}.\hskip 1em
  plus 0.5em minus 0.4em\relax Springer, 2015, pp. 67--101.

\bibitem{vershynin2010introduction}
R.~Vershynin, ``Introduction to the non-asymptotic analysis of random
  matrices,'' in \emph{Compressed Sensing: Theory and Applications}.\hskip 1em
  plus 0.5em minus 0.4em\relax Cambridge Univ. Press, 2012, pp. 210--268.

\bibitem{vershynin2018high}
------, \emph{High-dimensional probability: An introduction with applications
  in data science}.\hskip 1em plus 0.5em minus 0.4em\relax Cambridge University
  Press, 2018, vol.~47.

\bibitem{zou2005regularization}
H.~Zou and T.~Hastie, ``Regularization and variable selection via the elastic
  net,'' \emph{Journal of the royal statistical society: series B (statistical
  methodology)}, vol.~67, no.~2, pp. 301--320, 2005.

\bibitem{zou2006sparse}
H.~Zou, T.~Hastie, and R.~Tibshirani, ``Sparse principal component analysis,''
  \emph{Journal of computational and graphical statistics}, vol.~15, no.~2, pp.
  265--286, 2006.

\end{thebibliography}
\end{document}